\documentclass[12pt,reqno]{amsart}
\usepackage{amsmath}
\usepackage{amsthm}
\usepackage{amsfonts}
\usepackage{amssymb}
\usepackage[dvips]{graphicx}
\usepackage[flushleft]{caption2}
\newtheorem{theorem}{Theorem}

\newtheorem{lemma}[theorem]{Lemma}

\begin{document}

\title[Radiation for Difference Operators]%
{Radiation Conditions for the Difference Schr\"{o}dinger Operators
\vspace{.50in}}
\author[W.~Shaban \and B.~Vainberg]{W.~Shaban \and B.~Vainberg
\vspace{.25in}
\protect\\
Dept. of Mathematics\protect\\
UNC at Charlotte\protect\\
Charlotte, NC, 28223}
\thanks{The work of B. Vainberg was partly supported by the NSF Grant DMS-9971592.}


\date{}
\maketitle

\begin{abstract}
The problem of determining a unique solution of the
Schr\"{o}\-din\-ger equation $\left(  \Delta+q-\lambda\right)
\psi=f$ on the lattice $\mathbb{Z}^{d}$ is considered, where
$\Delta$ is the difference Laplacian and both $f$ and $q$ have
finite supports$.$ It is shown that there is an exceptional set
$S_{0}$ of points on $Sp(\Delta)=[-2d,2d]$ for which the limiting
absorption principle fails, even for unperturbed operator
($q(x)=0$). This exceptional set consists of the points $\left\{
\pm4n\right\}  $ when $d$ is even and $\left\{ \pm2(2n+1)\right\}
$ when $d$ is odd. For all values of $\lambda
\in[-2d,2d]\backslash S_{0},$ the radiation conditions are found
which single out the same solutions of the problem as the ones
determined by the limiting absorption principle. These solutions
are combinations of several waves propagating with different
frequencies, and the number of waves depends on the value of
$\lambda.$
\end{abstract}

\medskip\noindent\textbf{1991 AMS Classification:} 35P25, 47B39

\bigskip

\noindent\textbf{1. Introduction: }We investigate the problem of determining a
unique solution of the Schr\"{o}dinger equation
\begin{equation}
\left(  \Delta+q-\lambda\right)  \psi=f,\label{schro}%
\end{equation}
on the lattice $\mathbb{Z}^{d},$ where both $f$ and $q$ are functions with
bounded supports. Let $C_{0}(\mathbb{Z}^{d})$ be the set of such functions.

In the above equation, $\Delta$ is the difference Laplacian in $\mathbb{Z}%
^{d}$ defined by
\[
\Delta\psi\left(  \xi\right)  =\underset{|\xi^{\prime}-\xi|=1}{\sum}%
\psi\left(  \xi^{\prime}\right)  ,
\]
where $\xi=\left(  \xi_{1},\xi_{2},...,\xi_{d}\right)  \in\mathbb{Z}^{d}$\ and
$\psi\in l^{2}(\mathbb{Z}^{d}).$ The Fourier transform of $\psi$ is defined by
the formula
\[
\hat{\psi}\left(  k\right)  =\left(  2\pi\right)  ^{-d/2}\underset{\xi
\in\mathbb{Z}^{d}}{\sum}\psi\left(  \xi\right)  e^{-ik\cdot\xi},
\]
where $k=$ $\left(  k_{1},k_{2},...,k_{d}\right)  \in\left[  -\pi,\pi\right]
^{d},$ and $k\cdot\xi:=\overset{d}{\underset{i=1}{\sum}}k_{i}\xi_{i}$. The
Fourier transform of $\Delta$ is an operator of multiplication by $\phi\left(
k\right)  :$%
\begin{equation}
\widehat{\Delta}\widehat{\psi}(k)=\widehat{\Delta\psi}(k)=\phi\left(
k\right)  \widehat{\psi}(k),\quad\phi\left(  k\right)  =2\overset{d}%
{\underset{i=1}{\sum}}\cos k_{i}.\label{phi}%
\end{equation}
Thus, the operator $\Delta$ is self adjoint and its spectrum is a.c and
coincides with the range of the function $\phi,$ that is $Sp(\Delta)=\left[
-2d,2d\right]  .$ Hence, $Sp_{ess}\left(  \Delta+q\right)  =\left[
-2d,2d\right]  $ since $q$ has a bounded support$.$ \ As we shall see later,
there is an important exceptional set $S_{0}$ of values of $\lambda$ on the
interval $\left[  -2d,2d\right]  :$
\[
S_{0}:=\left\{  \pm4n\text{ when }d\text{ is even,}\pm2(2n+1)\text{ when
}d\text{ is odd, }n\in\mathbb{Z}\text{ and }2n\leq d\right\}  .
\]
Let
\[
S:=Sp(\Delta)\backslash S_{0}=\left[  -2d,2d\right]  \backslash S_{0}.
\]

There are two well known principles that are very natural from the point of
view of physics and which allow one to single out the unique solution of the
Schr\"{o}dinger equation in $\mathbb{R}^{d}$ (see \cite{ts}, \cite{v})$.$
These principles are the limiting absorption principle and the Sommerfeld
radiation conditions. It turns out that there is an essential difference in
both the validity and the form of these principles when applied to the
Schr\"{o}dinger equation in $\mathbb{R}^{d}$ and on the lattice. \ In
$\mathbb{R}^{d},$ the spectrum of the negative Laplacian is $Sp(-\Delta
)=\left\{  \lambda\geq0\right\}  .$ The limiting absorption principle and the
radiation conditions are valid for any $\lambda>0$ (see \cite{ts}, \cite{v}).
In \cite{es}, \cite{es2}, both of these principles are investigated for
general equations on the lattice of the form
\[
(A+q-\lambda)u=f,\quad q\in C_{0}(\mathbb{Z}^{d}\ ),
\]
where operator $A$ in the dual space (after the Fourier transform) is an
operator of multiplication by a smooth, real valued $2\pi-$periodic function
$a(k)$. \ In \cite{es}, \cite{es2}, the limiting absorption principle is
justified for values of $\lambda$ such that $\nabla a(k)\neq0$ on the surface
\[
\Gamma(\lambda)=\{k:k\in T^{d},\text{ }a(k)=\lambda\},
\]
and the radiation conditions are found when this surface is strictly convex.
Here $T^{d}$ is the torus $\mathbb{R}^{d}/2\pi\mathbb{Z}^{d}$. \

By applying results from \cite{es}, \cite{es2} to the difference
Schr\"{o}dinger equation (\ref{schro}), one gets the limiting absorption
principle when $\lambda\in S,$ and the radiation conditions when $\lambda$
belongs to the following two intervals on the continuous spectrum of the
Laplacian:
\[
2d-4<\left|  \lambda\right|  <2d.
\]
The goal of this paper is to investigate the validity of the
limiting absorption principle when $\lambda\in S_{0}$ and, more
importantly, find the radiation conditions when $\left|
\lambda\right|  <2d-4$. \ Possibly, one of the most important
observations in this paper consists of the fact that the
fundamental solutions of the difference Schr\"{o}dinger operator
with $\left| \lambda\right|  <2d-4,$ singled out by the limiting
absorption principle or by the radiation conditions, decay at
infinity as fast as in the continuous case only for non-singular
directions. The decay is much slower for singular directions. It
is also important that the asymptotic behavior of the fundamental
solutions at infinity is more complicated, and it changes
dramatically when $\lambda$ passes through the points of the set
$S_{0}$ . \

Let us recall that the limiting absorption principle enables us to obtain two
solutions $\psi_{\pm}$ of (\ref{schro}), for any $\lambda\in S,$ as a
(pointwise) limit of $\psi_{_{\eta}}\left(  \xi\right)  =\left(
\mathcal{R}_{_{\eta}}f\right)  \left(  \xi\right)  $ as $\eta\rightarrow
\lambda\pm i0.$ Here,
\[
\mathcal{R}_{_{\eta}}=\left(  \Delta+q-\eta\right)  ^{-1}:l^{2}\left(
\mathbb{Z}^{d}\right)  \longrightarrow l^{2}\left(  \mathbb{Z}^{d}\right)
,\quad\operatorname{Im}\eta\neq0.
\]
In order to clarify the situation with $\lambda\in S_{0},$ one can consider
the simplest case $d=2$ where the set $S_{0}$ consists of two end points
$\lambda=\pm4$ of the spectrum and the point $\lambda=0.$ It can be shown that
the exponentially decaying at infinity fundamental solution $E_{_{\eta}}(\xi)$
of the operator $\Delta-\eta,$ $\xi\in\mathbb{Z}^{2},$ $\operatorname{Im}%
\eta>0,$ has the following form as $\eta\rightarrow\pm i0:$%
\[
E_{_{\eta}}(\xi)=\pm C(\xi)\ln\left|  \eta\right|  +\widetilde{E}_{\pm}%
(\xi)+o(1),\quad C(\xi)=\frac i{8\pi}\left[  \left(  -1\right)  ^{\xi_{1}%
}+\left(  -1\right)  ^{\xi_{2}}\right]  ,
\]
Thus, $E_{_{\eta}}$ does not have a (pointwise) limit as $\eta\rightarrow\pm
i0,$ and therefore the limiting absorption principle is not valid when
$\lambda=0.$ \ Note that $C(\xi)$\ satisfies the equation $\Delta C(\xi)=0,
$\ and therefore $\widetilde{E}_{\pm}$ are fundamental solutions of the
Laplacian. However, these fundamental solutions grow logarithmically at
infinity, unlike the fundamental solutions obtained by the limiting absorption
principle. The latter ones decay as $\left|  \xi\right|  ^{-1/2}$ in the case
of $d=2.$ The statements above and similar statements when $d>2$ are not very
difficult to prove, but the proofs are rather technical, and we do not include
them in this paper.

The other way to single out unique solutions of (\ref{schro}) is by imposing
some conditions at infinity called the Sommerfeld radiation conditions, or by
requiring a special asymptotic behavior of the solution at infinity. We find
the asymptotic behavior of solutions and the radiation conditions for the
difference Schr\"{o}dinger operators with arbitrary $\lambda\in S.$ We show
that, for any fixed $\lambda\in S,$ the radiation conditions or the asymptotic
behavior of the solutions single out the same two solutions of (\ref{schro})
found by the limiting absorption principle. One of the two solutions
corresponds to waves propagating to infinity and the other one corresponds to
waves coming from infinity. These waves are not spherical as in the case of
$\mathbb{R}^{d}$ . More importantly, in the lattice case, each solution of
(\ref{schro}) is a combination of several waves propagating with different
frequencies, and the number of waves depends on the value of $\lambda.$ (Only
one wave exists if $2d-4<\left|  \lambda\right|  <2d).$ For equations in
$\mathbb{R}^{d},$ several different waves appear in the case of nonisotropic
elasticity equations with compactly supported right-hand sides and in the case
of more general systems of equations of higher order (see \cite{v}), but not
for the Schr\"{o}dinger equation. In fact, our approach to study the
difference equation (\ref{schro}) follows the one used in \cite{v} to study
the general systems in $\mathbb{R}^{d}.$

The form of the radiation conditions and the asymptotic behavior of the
solutions of (\ref{schro}) depends dramatically on the shape of the surface
$\Gamma(\lambda),$%

\begin{equation}
\Gamma(\lambda)=\{k:k\in T^{d},\text{ }\phi\left(  k\right)  =\lambda
\},\quad\lambda\in\left[  -2d,2d\right]  .\label{star101}%
\end{equation}
Here $T^{d}=\mathbb{R}^{d}/2\pi\mathbb{Z}^{d},$ function $\phi$ is
defined in (\ref{phi}). \ We fix the orientation of $\Gamma
(\lambda )$ by choosing the normal vector
\[
n:=\nabla \phi \left( k\right) .
\]

If $2d-4<\left|  \lambda\right|  <2d,$ then the surface
$\Gamma(\lambda)$ is strictly convex, and there is a unique point
\[
k=k\left(  \omega,\lambda\right)  \in\Gamma(\lambda)
\]
where the normal $n$ to the surface $\Gamma(\lambda)$ is parallel
to and has the same direction as the unit vector
$\omega=\frac{\xi}{\left|  \xi\right|  }.$ In the remaining part
of $S=Sp(\Delta)\backslash S_{0}$ when $\left|  \lambda\right|
<2d-4,$ the surface $\Gamma(\lambda)$ is not convex and, for some
$\omega\in\Omega,$ there exist more than one point at which the
normal $n$ to the surface is parallel to and has the same
direction as $\omega$ (see Fig $4,5$ below)$.$ In addition, the
curvature of the surface $\Gamma(\lambda)$ is zero at some points.

We shall call a point $\omega\in\Omega$ singular if there is a
point $k(\omega,\lambda)$ on the surface $\Gamma(\lambda)$ at
which the normal $n$ to this surface is parallel to $\omega$ and
the total curvature (the product of principle curvatures) of the
surface at $k(\omega,\lambda)$ is zero. For a fixed $\lambda\in
S,$ let $\Omega_{0}$ be the set of singular points $\omega$ of
$\Omega.$ \ The set $\Omega \backslash\Omega_{0}$ is an open
subset of $\Omega$ and consists of a finite number of connected
components. We shall call these components non-singular domains.
For any non-singular domain $V\subset\Omega$, let $m_{V}$ be the
number of the points $k(\omega,\lambda;s),$ $1\leq s\leq m_{V},$
at which the normal $n=\nabla\phi\left(  k\right)  $ to the
surface $\Gamma(\lambda)$ is parallel to and has the same
direction as $\omega$. Let $\sigma =\sigma (V,s)$ be the
difference between the number of positive and negative principle
curvatures at the point $k(\omega ,\lambda ;s).$ Since $V$ is
connected and the multi valued function $\omega \rightarrow
\{k(\omega ,\lambda ;s)\},$ where $\lambda $ is fixed,
$\omega \in V,$ is smooth, then $%
m_{V}$ and $\sigma $ do not depend on $\omega \in V.$ They depend
only on the domain $V.$ One can show that $m_{V}\leq2^{d}d$. \ Let
$\mu(\omega ,\lambda;s)=k(\omega,\lambda;s)\cdot\omega$ be the
projection of the vector $k(\omega,\lambda;s)$ on $\omega.$ We
denote the cube $[-r,r]^{d}$ in $\mathbb{Z}^{d}$ by $B_{r}.$

We prove that, for any $\lambda\in S$, equation (\ref{schro}) admits unique
solutions $\psi_{+}$ and $\psi_{-}$ such that for any integer $R\geq1,$%
\begin{equation}
\frac1R\underset{\xi\in B_{2R}\backslash B_{R}}{\sum}\left|  \psi_{\pm}%
(\xi)\right|  ^{2}<C<\infty,\label{rads1}%
\end{equation}
and for any non-singular domain $V\subset\Omega,$ and $\omega=\frac\xi{\left|
\xi\right|  }\in V,$
\begin{equation}
\psi_{\pm}\left(  \xi\right)  =\text{ }\overset{m_{V}}{\underset{s=1}{\sum}%
}\frac{e^{\pm i\mu\left(  \omega,\lambda;s\right)  \left|  \xi\right|  }%
}{\left|  \xi\right|  ^{\frac{d-1}2}}a_{\pm}\left(  \omega,\lambda;s\right)
+O\left(  \frac1{\left|  \xi\right|  ^{\frac{d+1}2}}\right)  \quad\text{as
}\left|  \xi\right|  \longrightarrow\infty,\label{asym}%
\end{equation}
where the coefficients $a_{\pm}\left( \omega,\lambda;s\right)  $
are smooth and the remainder can be estimated uniformly in
$\omega$ on any compact set $Q\subset V,$ . Moreover, these
solutions are equal to $R_{\lambda\pm i0}f.$ The same solutions
can be singled out by the following radiation conditions at
infinity: \ instead of (\ref{asym}), one can assume that
$\psi_{\pm}$ can be represented as a sum
\begin{equation}
\psi_{\pm}(\xi)=\overset{m_{V}}{\underset{s=1}{\sum}}\psi_{s\pm}(\xi
),\quad\omega=\frac\xi{\left|  \xi\right|  }\in V,\label{34'}%
\end{equation}
where each term $\psi_{s\pm}(\xi)$ satisfies the following conditions as
$\left|  \xi\right|  \longrightarrow\infty,$
\begin{equation}
\left\{
\begin{array}
[c]{l}%
\psi_{s\pm}\left(  \xi\right)  =O\left(  \frac1{\left|  \xi\right|
^{\frac{d-1}2}}\right) \\
\psi_{s\pm}\left(  \xi+e_{j}\right)  =e^{\pm ik_{j}\left(  \omega
,\lambda;s\right)  }\psi_{s\pm}\left(  \xi\right)  +O\left(  \frac1{\left|
\xi\right|  ^{\frac{d+1}2}}\right)  ,\quad\text{ }j=1,...,d,
\end{array}
\right. \label{rads2}%
\end{equation}
with the remaining terms decaying uniformly in $\omega=\frac\xi{\left|
\xi\right|  }$ $\in Q$ for any compact set $Q\subset V.$ Here $e_{j}=\left(
0,...,0,\underset{jth}{1},0,...,0\right)  $ and $k_{j}\left(  \omega
,\lambda;s\right)  $ is the $j-$th coordinate of the point $k\left(
\omega,\lambda;s\right)  .$ \ Note that the second condition in (\ref{rads2})
can be written in the following form which is more similar to the standard
radiation condition in $\mathbb{R}^{d}$:
\[
\frac{\partial\psi_{s\pm}}{\partial\xi_{j}}\left(  \xi\right)  =\left(  e^{\pm
ik_{j}\left(  \omega,\lambda;s\right)  }-1\right)  \psi_{\pm}\left(
\xi\right)  +O\left(  \left|  \xi\right|  ^{-\frac{d+1}2}\right)  ,
\]
where
\[
\frac{\partial\psi}{\partial\xi_{j}}\left(  \xi\right)  :=\psi\left(
\xi+e_{j}\right)  -\psi\left(  \xi\right)  .
\]

Let us note again that $\Omega_{0}$ is empty when $2d-4<\left|  \lambda
\right|  <2d.$ \ In this case,
\begin{equation}
\psi_{\pm}=O\left(  \left|  \xi\right|  ^{\frac{1-d}{2}}\right)  ,\quad\left|
\xi\right|  \rightarrow\infty,\label{az}%
\end{equation}
and this estimate is uniform in $\omega\in\Omega.$ \ Estimate (\ref{az}) fails
when $\left|  \lambda\right|  <2d-4$ and $\omega\in\Omega_{0}.$

The plan of the paper is as follows. In section 2, we discuss the limiting
absorption principle and the radiation conditions for the unperturbed equation
($q=0).$ In section 3, we carry out the obtained results to the Schr\"{o}%
dinger equation. The appendix contains the proofs of the lemmas on the
properties of the surface $\Gamma\left(  \lambda\right)  .\bigskip$

\noindent\textbf{2. The Unperturbed Problem: }We start with an investigation
of the unperturbed problem (\ref{schro})$:$\
\begin{equation}
\left(  \Delta-\lambda\right)  \psi=f,\quad\xi\in\mathbb{Z}^{d},\label{lap}%
\end{equation}
which describes the propagation of waves in a homogeneous medium$.$ When
$\lambda\notin\lbrack-2d,$ $2d]$, the resolvent of the difference Laplacian is
a bounded operator in $l^{2}(\mathbb{Z}^{d}),$ and it is given by the formula
\begin{equation}
\left(  R_{_{\lambda}}f\right)  \left(  \xi\right)  =\frac{1}{\left(
2\pi\right)  ^{d/2}}\int_{T^{d}}\frac{\widehat{f}\left(  k\right)
e^{ik\cdot\xi}}{\phi\left(  k\right)  -\lambda}dk,\quad\text{where }%
T^{d}=\mathbb{R}^{d}/2\pi\mathbb{Z}^{d},\label{res}%
\end{equation}
In (\ref{res}) and in other similar formulas, we shall always identify $T^{d}
$ with the cube $\left[  -\pi,\pi\right]  ^{d}$ if $\lambda>0,$ and with the
cube $\left[  0,2\pi\right]  ^{d}$ if $\lambda<0.$ \ If $\lambda\in S,$ then
$|\nabla\phi|$ $\neq$ $0$ on the surface $\Gamma(\lambda).$ This allows to
prove (see \cite{es}, \cite{v}) the following statement.

\begin{theorem}
\label{at}For any $\lambda\in S,$ any $f\in C_{0}(\mathbb{Z}^{d}),$ and any
fixed $\xi\in\mathbb{Z}^{d},$ the pointwise limits of $R_{_{\eta}}f$ as
$\eta\rightarrow\lambda\pm i0$ exist and are given by the following
expression:
\begin{align}
(R_{\lambda\pm i0}f)\left(  \xi\right)   & =\frac1{\left(  2\pi\right)
^{d/2}}\int_{T^{d}}\frac{(1-\chi\left(  k\right)  )\widehat{f}\left(
k\right)  e^{ik\cdot\xi}}{\phi\left(  k\right)  -\lambda}dk\nonumber\\
& +\frac1{(2\pi)^{d/2}}\oint_{\left|  \rho-\lambda\right|  <\delta}%
\int_{\Gamma(\rho)}\frac{\chi\left(  k\right)  \widehat{f}\left(  k\right)
e^{ik\cdot\xi}}{\rho-\lambda}\frac1{|\nabla\phi|}dsd\rho\nonumber\\
& \pm\frac{\pi i}{(2\pi)^{d/2}}\int_{\Gamma(\lambda)}\frac{\widehat{f}\left(
k\right)  e^{ik\cdot\xi}}{|\nabla\phi|}ds,\label{ResL2}%
\end{align}
where $\chi$ is an infinitely smooth function on $T^{d}$ such that
\[
\chi\left(  k\right)  =\left\{
\begin{array}
[c]{c}%
0\qquad\text{if }\left|  \phi\left(  k\right)  -\lambda\right|  >\delta
\hfill\\
1\qquad\text{if }\left|  \phi\left(  k\right)  -\lambda\right|  \leq\delta/2,
\end{array}
\right.
\]
with $0<\delta<dist(\lambda,S_{0}),$ the surface $\Gamma(\rho)$ is defined in
(\ref{star101}), and $ds$ is the surface element.
\end{theorem}

In order to describe the asymptotic behavior at infinity of the solutions to
the equation (\ref{lap}) constructed in Theorem \ref{at} we need to know
geometrical properties of the surface $\Gamma(\lambda).$ The following
statement is proved in the appendix:

\begin{lemma}
\label{convexity}When $2d-4<\lambda<2d$ $\left(  -2d<\lambda<-2d+4\right)  ,$
the surface $\Gamma\left(  \lambda\right)  $ is located strictly inside the
cube $\left[  -\pi,\pi\right]  ^{d}$ $($ $\left[  0,2\pi\right]  ^{d},$
respectively)$,$ and it is smooth, convex and closed with the curvature $K(k)
$ not vanishing at any point$.$
\end{lemma}

Graphs of the surface $\phi\left(  k\right)  =\lambda,$
$k\in\mathbb{R}^{d},$ are shown in Figures 1-3 for
$\lambda=0,1,-1,$ and $d=2.$ The surface $\Gamma\left(
\lambda\right)  $ can be obtained if these graphs are taken by
modulo $\ 2\pi\mathbb{Z}^{2}.$ As it was mentioned at the
beginning of this section, we identify $\Gamma\left(
\lambda\right)  $ with the graph of $\phi\left( k\right) =\lambda$
in the cube $\left[  -\pi,\pi\right]  ^{d}$ if $\lambda>0,$ or in
the cube $\left[  0,2\pi\right]  ^{d}$ if $\lambda<0.$ Figures 4
and 5 below give the graphs of $\Gamma(\lambda)$ for $\lambda=1$
and $\lambda=3$
respectively$,$ and $d=3.$%

\begin{figure}[ht]
\centering
\begin{minipage}[b]{0.49\textwidth}
\centering
\includegraphics[width=2.4in]{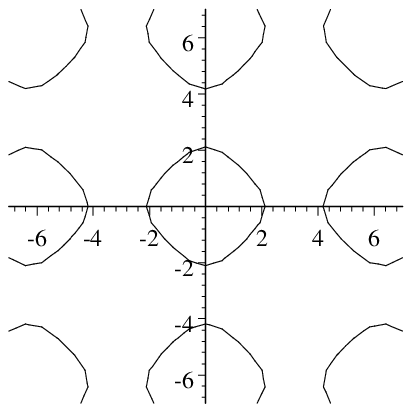}
\par\vspace{0pt}
\caption{$d=2,\lambda=1$}
\end{minipage}
\hspace{\fill}
\begin{minipage}[b]{0.49\textwidth}
\centering
\includegraphics[width=2.4in]{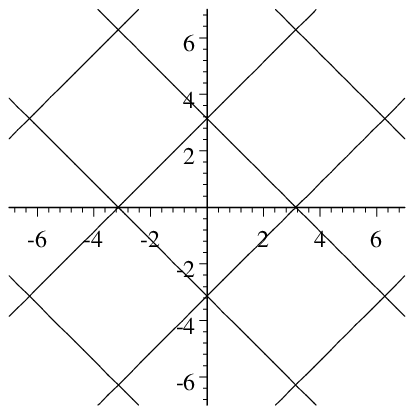}
\par\vspace{0pt}
\caption{$d=2, \lambda=0$}
\end{minipage}
  \begin{minipage}[b]{0.49\textwidth}
    \centering
    \includegraphics[width=2.4in]{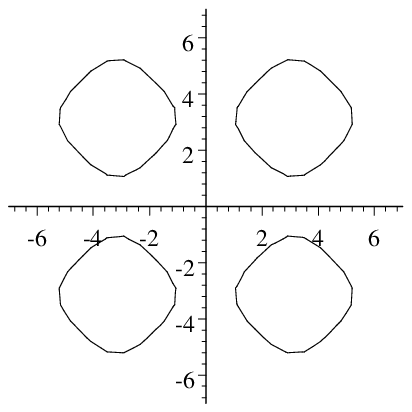}
    \par\vspace{0pt}
    \caption{$d=2,\lambda=-1$}
  \end{minipage}
\end{figure}

We specify the orientation on the surface $\Gamma\left(  \lambda\right)  $ by
choosing the normal vector $n$ to be
\[
n=\nabla\phi\left(  k\right)  =-2\left(  \sin k_{1},...,\sin k_{d}\right)  .
\]
When $\left|  \lambda\right|  <2d-4,$ the surface $\Gamma(\lambda)$ is not
convex and may have several normal vectors with the same direction. The
following statement is also proved in the appendix:

\begin{lemma}
\label{normals}The number $m$ of points of $\Gamma(\lambda),\lambda\in S,$ at
which the normal $n=\nabla\phi$ to $\Gamma(\lambda)$ is parallel to and has
the same direction as a fixed unit vector $\omega,$ and the total curvature of
the surface at these points is non-zero, is such that $m\leq2^{d}d.$

( Note that the above inequality gives a very rough estimate. For example,
$m\leq4$ when $d=3.)$
\end{lemma}

We denote by $W_{\pm}$ the two classes of functions $\psi_{\pm}(\xi)$ for
which (\ref{rads1}) and (\ref{asym}) hold. \ Let $W_{\pm}^{\prime}$ be two
classes of functions $\psi_{\pm}(\xi)$ that satisfy the estimate $\left(
\ref{rads1}\right)  $ and can be represented in the form (\ref{34'}),
(\ref{rads2}).

\begin{figure}[ht]
\centering
\begin{minipage}[b]{0.49\textwidth}
\centering
\includegraphics[width=2.7in]{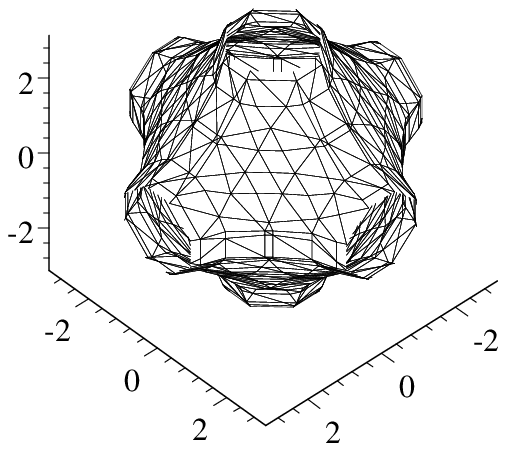}
\par\vspace{0pt}
\caption{$d=3,\lambda=1$}
\end{minipage}
\hspace{\fill}
\begin{minipage}[b]{0.49\textwidth}
\centering
\includegraphics[width=2.7in]{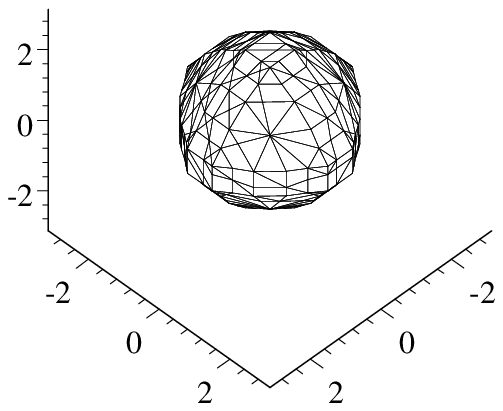}
\par\vspace{0pt}
\caption{$d=3, \lambda=3$}
\end{minipage}
\end{figure}

\begin{theorem}
\label{asym11}For any $f\in C_{0}\left(  \mathbb{Z}^{d}\right)  $and for any
$\lambda\in S,$ the equation
\begin{equation}
\left(  \Delta-\lambda\right)  \psi=f\label{door2}%
\end{equation}
admits unique solutions in the classes $W_{+},W_{-}$ and unique solutions in
$W_{+}^{\prime},W_{-}^{\prime}.$ The solutions in $W_{\pm}$ and the
corresponding ones in $W_{\pm}^{\prime}$ are the same and coincide with
$R_{\lambda\pm i0}f$.
\end{theorem}

\noindent\textbf{Remark }The amplitudes $a_{+}$\ and $a_{-}$\ in expansion
(\ref{asym})\ are equal to $a$\ and $\overline{a}$ respectively,
where\textit{\ }$\overline{a}$ is the complex conjugate of $a$ \textit{\ }and
\begin{equation}
a=a\left(  \omega,\lambda;s\right)  =\frac{\sqrt{2\pi}\widehat{f}\left(
k\left(  \omega,\lambda;s\right)  \right)  e^{i(\sigma +2)\frac{\pi}{4}}}%
{\sqrt{\left|  K(\omega,\lambda;s))\right|  }\left|  \nabla\phi(k\left(
\omega,\lambda;s\right)  )\right|  }.\label{amp1}%
\end{equation}
Here $K(\omega,\lambda;s))$ is the total curvature of the surface
$\Gamma(\lambda)$ at the point $k\left(  \omega,\lambda;s\right),
$ $\sigma =\sigma (V,s)$ is the difference between the number of
positive and negative principle curvatures at $k\left( \omega
,\lambda ;s\right) .$

In order to prove this theorem, we need the following lemma.

\begin{lemma}
\label{lem11}Let $\Gamma\subset T^{d}$ be a smooth surface defined by the
equation
\[
k_{j}=g\left(  k^{\prime}\right)  ,\quad k^{\prime}=(k_{1},...k_{j-1}%
,k_{j+1},...,k_{d})\in T^{d-1},
\]
and let
\[
F\left(  \xi\right)  =\left(  2\pi\right)  ^{1-\frac d2}\int_{\Gamma}f\left(
k\right)  e^{ik\cdot\xi}dk^{\prime},\quad\xi\in\mathbb{Z}^{d},\quad f\in
C^{\infty}\left(  T^{d}\right)  .
\]
Then for any $R>0,$%
\begin{equation}
\underset{\xi\in B_{R}}{\sum}\left|  F\left(  \xi\right)  \right|  ^{2}%
\leq2\pi\left(  2R+1\right)  \int_{\Gamma}\left|  f\left(  k\right)  \right|
^{2}dk^{\prime}.\label{new100}%
\end{equation}
\end{lemma}%

\proof
For each fixed $\xi_{j},$ the function $F$ is the Fourier
transform of the function
\[
h\left(  k^{\prime}\right)  =\sqrt{2\pi}f\left(  k\right)  e^{ik_{j}\xi_{j}%
}\left|  _{k_{j}=g\left(  k^{\prime}\right)  }\right.  .
\]
Hence, from Parseval's equality it follows that%

\[
\underset{\xi^{\prime}\in\mathbb{Z}^{d-1}}{\sum}\left|  F\left(  \xi^{\prime
},\xi_{d}\right)  \right|  ^{2}=\int_{T^{d-1}}\left|  h\left(  k^{\prime
}\right)  \right|  ^{2}dk^{\prime}=2\pi\int_{\Gamma}\left|  f\left(  k\right)
\right|  ^{2}dk^{\prime},
\]
which immediately implies (\ref{new100}).%
\endproof
\medskip

\noindent\textbf{Proof of Theorem \ref{asym11}.} To prove the theorem, we show
first that (\ref{asym}) implies (\ref{rads2})$,$ and therefore $W_{\pm}\subset
W_{\pm}^{\prime}$. Then, we show that $R_{\lambda\pm i0}f$ are the solutions
of (\ref{lap}) which belong to $W_{\pm}.$ After this, it suffices to prove
only the uniqueness of the solutions in $W_{\pm}^{\prime} $.

Let $\psi_{\pm}$ satisfy (\ref{asym}). Then, obviously, $\psi_{\pm}$ can be
represented in the form (\ref{34'}) with
\begin{equation}
\psi_{s\pm}\left(  \xi\right)  =\text{ }\frac{e^{\pm i\mu\left(
\omega,\lambda;s\right)  \left|  \xi\right|  }}{\left|  \xi\right|
^{\frac{d-1}{2}}}a_{\pm}\left(  \omega,\lambda;s\right)  +O\left(  \left|
\xi\right|  ^{-\frac{d+1}{2}}\right)  ,\quad\text{ }\left|  \xi\right|
\longrightarrow\infty.\label{V}%
\end{equation}
Thus, the first relation of (\ref{rads2}) holds, and
\[
\psi_{s\pm}\left(  \xi+e_{j}\right)  =\text{ }\frac{e^{\pm ik\left(
\omega^{\prime},\lambda;s\right)  \cdot\left(  \xi+e_{j}\right)  }}{\left|
\xi+e_{j}\right|  ^{\frac{d-1}{2}}}a_{\pm}\left(  \omega^{\prime}%
,\lambda;s\right)  +O\left(  \left|  \xi\right|  ^{-\frac{d+1}{2}}\right)
\quad\text{as }\left|  \xi\right|  \longrightarrow\infty,
\]
where $\omega^{\prime}=\frac{\xi+e_{j}}{\left|  \xi+e_{j}\right|  }$. By
taking into account that
\[
\left|  \xi+e_{j}\right|  =\left|  \xi\right|  \left(  1+O\left(  \left|
\xi\right|  ^{-1}\right)  \right)  ,\text{\quad}\omega^{\prime}=\omega
+O\left(  \left|  \xi\right|  ^{-1}\right)  ,
\]
and $\nabla_{_{\omega}}k\left(  \omega,\lambda\right)  $ is orthogonal to
$\omega,$ we obtain that
\[
\psi_{s\pm}\left(  \xi+e_{j}\right)  =\text{ }\frac{e^{\pm ik\left(
\omega,\lambda;s\right)  \cdot\left(  \xi+e_{j}\right)  }}{\left|  \xi\right|
^{\frac{d-1}{2}}}a_{\pm}\left(  \omega,\lambda;s\right)  +O\left(  \left|
\xi\right|  ^{-\frac{d+1}{2}}\right)  \quad\text{as }\left|  \xi\right|
\longrightarrow\infty,
\]
which together with (\ref{V}) immediately leads to the second part of
(\ref{rads2})$.$ Thus, the asymptotic behavior (\ref{asym}) implies the
radiation conditions (\ref{rads2})$,$ and the inclusion $W_{\pm}\subset
W_{\pm}^{\prime}$ is proved.

Next, we prove that $\psi_{\pm}=R_{\lambda\pm i0}f$ are solutions of
(\ref{lap}) and $\psi_{\pm}\in W_{\pm}$. First, let us note that $R_{_{\eta}%
}f,$ defined in (\ref{ResL2}), satisfies (\ref{lap}) with $\lambda=\eta
\notin\left[  -2d,2d\right]  ,$ and from Theorem \ref{at} it follows that one
can pass to the limit in the equation as $\eta\rightarrow\lambda\pm i0.$ Thus,
$\psi_{\pm}$ are solutions of (\ref{lap})$.$ Let us show that $\psi_{\pm}$
satisfy (\ref{asym})$.$

The first integrand on the right-hand side of (\ref{ResL2}) is a smooth
function because $\chi\left(  k\right)  =1$ whenever $\phi\left(  k\right)
=\lambda.$ By integrating by parts as many times as needed, we obtain that the
first term on the right-hand side of (\ref{ResL2}) has order $O\left(  \left|
\xi\right|  ^{-\infty}\right)  $ as $\left|  \xi\right|  \longrightarrow
\infty.$ Thus, (\ref{ResL2}) implies
\begin{equation}
\psi_{\pm}\left(  \xi\right)  =\frac1{2\pi}\int_{\left|  \rho-\lambda\right|
<\delta}\frac{\Phi\left(  \xi,\rho\right)  }{\rho-\lambda}d\rho\text{ }%
\pm\frac i2\Phi\left(  \xi,\lambda\right)  +O\left(  \left|  \xi\right|
^{-\infty}\right)  ,\quad\left|  \xi\right|  \rightarrow\infty.\label{4th}%
\end{equation}
where
\begin{equation}
\Phi\left(  \xi,\rho\right)  :=(2\pi)^{1-\frac d2}\int_{\Gamma(\rho)}%
\frac{\chi\left(  k\right)  \widehat{f}\left(  k\right)  e^{ik\cdot\xi}%
}{|\nabla\phi|}ds.\text{ }\label{411th}%
\end{equation}

We apply the stationary phase method to the integral (\ref{411th}) in order to
get the asymptotic behavior of $\Phi\left(  \xi,\rho\right)  $ as $\left|
\xi\right|  \longrightarrow\infty$. The asymptotic behavior of $\Phi\left(
\xi,\rho\right)  $ depends on points $k\left(  \omega,\rho\right)  $ on the
surface $\Gamma(\rho)$ at which the normal $n=\nabla\phi\left(  k\right)  $ to
the surface is parallel to $\omega=\frac\xi{\left|  \xi\right|  }$ (see
\cite{v}, Theorem 9 of Chapter I ). If $\rho=\lambda$ and $\omega$ belongs to
a nonsingular domain $V\subset\Omega$ then there are $m_{V}$ points $k\left(
\omega,\lambda;s\right)  $ on $\Gamma(\lambda)$\ at which the normal
$n=\nabla\phi\left(  k\right)  $ to the surface $\Gamma(\lambda)$ is parallel
to and has the same direction as $\omega.$ Due to the symmetry of
$\Gamma(\lambda),$\ there are exactly $m_{V}$ points $k=-k\left(
\omega,\lambda;s\right)  $ where the directions of $n$ and $\omega$ are
opposite. The total curvature of $\Gamma(\lambda)$ at points $\pm k\left(
\omega,\lambda;s\right)  $ is not vanishing. Then from (\cite{v}, Theorem 9 of
Chapter I) it follows that
\begin{align}
\Phi\left(  \xi,\lambda\right)   & =-i\overset{m_{V}}{\underset{s=1}{\sum}%
}\left(  e^{i\mu\left(  \omega,\lambda;s\right)  \left|  \xi\right|  }a\left(
\omega,\lambda;s\right)  -e^{-i\mu\left(  \omega,\lambda;s\right)  \left|
\xi\right|  }\overline{a\left(  \omega,\lambda;s\right)  }\right)  \left|
\xi\right|  ^{-\frac{d-1}2}\nonumber\\
& +O\left(  \left|  \xi\right|  ^{-\frac{d+1}2}\right)  ,\quad\omega=\frac
\xi{\left|  \xi\right|  }\in V,\quad\left|  \xi\right|  \rightarrow
\infty,\label{117*}%
\end{align}
where $a\left(  \omega,\lambda;s\right)  $ is defined in (\ref{amp1}) and the
expansion is uniform with respect to $\omega$ in any compact subset of $V$.

Let us fix a compact $Q\subset V.$ The surface $\Gamma(\rho)$ depends
analytically on $\rho$ when $\lambda\in S$ and $\left|  \rho-\lambda\right|  $
is small enough$.$ Thus, if $\omega=\frac\xi{\left|  \xi\right|  }\in Q,$
$\left|  \rho-\lambda\right|  \leq\delta_{1}$ and $\delta_{1}$ \ is small
enough, then: 1) there are exactly $m_{V}$ points $k\left(  \omega
,\rho;s\right)  $ on $\Gamma(\rho)$\ at which the normal $n=\nabla\phi\left(
k\right)  $ to the surface $\Gamma(\rho)$ is parallel to and has the same
direction as $\omega,$ 2) the points depend analytically on $\rho$ and 3) the
total curvature of $\Gamma(\rho)$ at the points $\pm k\left(  \omega
,\rho;s\right)  $ is not vanishing$.$ Finally (see Remark 2 after Theorem 9 of
Chapter I in \cite{v}), expansion (\ref{117*}) is valid with $\lambda$
replaced by $\rho$ when $\omega\in Q,$ $\left|  \rho-\lambda\right|
\leq\delta_{1}:$
\begin{align}
\Phi\left(  \xi,\rho\right)   & =-i\overset{m_{V}}{\underset{s=1}{\sum}%
}\left(  e^{i\mu\left(  \omega,\rho;s\right)  \left|  \xi\right|  }a\left(
\omega,\rho;s\right)  -e^{-i\mu\left(  \omega,\rho;s\right)  \left|
\xi\right|  }\overline{a\left(  \omega,\rho;s\right)  }\right)  \left|
\xi\right|  ^{-\frac{d-1}2}\nonumber\\
& +O\left(  \left|  \xi\right|  ^{-\frac{d+1}2}\right)  ,\label{17*a}%
\end{align}
and it is uniform and admits differentiation with respect to all arguments.

We shall choose the constant $\delta$ in (\ref{ResL2}) and (\ref{4th}) so
small that $\delta<\delta_{1}.$ Then (\ref{4th}), (\ref{17*a}), and the
following simple estimate
\[
\int_{\left|  \rho-\lambda\right|  <\delta}\frac{F\left(  \rho\right)  }%
{\rho-\lambda}d\rho\leq2\delta\underset{\left|  \rho-\lambda\right|
\leq\delta}{\max}\left|  F^{\prime}\left(  \rho\right)  \right|
\]
which holds for any differentiable function $F$, imply that $\left|
\xi\right|  ^{\frac{d-1}{2}}\psi_{\pm}\left(  \xi\right)  $ is equal to
\begin{align*}
& \frac{-i}{2\pi}\int_{\left|  \rho-\lambda\right|  <\delta}\frac{1}%
{\rho-\lambda}\overset{m_{V}}{\underset{s=1}{\sum}}\left(  e^{i\mu\left(
\omega,\rho;s\right)  \left|  \xi\right|  }a\left(  \omega,\rho;s\right)
-e^{-i\mu\left(  \omega,\rho;s\right)  \left|  \xi\right|  }\overline{a\left(
\omega,\rho;s\right)  }\right)  d\rho\\
& \pm\frac{1}{2}\overset{m_{V}}{\underset{s=1}{\sum}}\left(  e^{i\mu\left(
\omega,\lambda;s\right)  \left|  \xi\right|  }a\left(  \omega,\lambda
;s\right)  -e^{-i\mu\left(  \omega,\lambda;s\right)  \left|  \xi\right|
}\overline{a\left(  \omega,\lambda;s\right)  }\right)  +O\left(  \left|
\xi\right|  ^{-1}\right)  ,
\end{align*}
when $\omega\in Q,$ $\left|  \xi\right|  \rightarrow\infty.$ \ Now from Lemma
5 in (\cite{v}, Chapter VII) on asymptotic behavior of the principal value
integrals it follows that
\begin{align}
\psi_{\pm}\left(  \xi\right)   & =\frac{1}{2}\overset{m_{V}}{\underset
{s=1}{\sum}}\gamma_{s}\left(  e^{i\mu\left(  \omega,\lambda;s\right)  \left|
\xi\right|  }a\left(  \omega,\lambda;s\right)  +e^{-i\mu\left(  \omega
,\lambda;s\right)  \left|  \xi\right|  }\overline{a\left(  \omega
,\lambda;s\right)  }\right)  \left|  \xi\right|  ^{-\frac{d-1}{2}}\nonumber\\
& \pm\frac{1}{2}\text{
}\overset{m_{V}}{\underset{s=1}{\sum}}\left( e^{i\mu\left(
\omega,\lambda;s\right)  \left|  \xi\right| }a\left(
\omega,\lambda;s\right)  -e^{-i\mu\left( \omega,\lambda;s\right)
\left| \xi\right|  }\overline{a\left( \omega,\lambda;s\right)
}\right)  \left|
\xi\right|  ^{-\frac{d-1}{2}}\nonumber\\
& +O\left(  \left|  \xi\right|  ^{-\frac{d+1}{2}}\right)  ,\text{ }\quad
\gamma_{s}=\text{sign}\left.  \frac{\partial}{\partial\rho}\left(  \mu\left(
\omega,\rho;s\right)  \right)  \right|  _{\rho=\lambda},\label{newasym1}%
\end{align}
where $\omega\in Q,$ $\left|  \xi\right|  \rightarrow\infty.$ \ In order to
find $\gamma_{s},$ we note that $\phi(k\left(  \omega,\rho;s\right)  )=\rho,$
since $k\left(  \omega,\rho;s\right)  \in$ $\Gamma(\rho).$ Thus,
\begin{equation}
\nabla\phi(k\left(  \omega,\rho;s\right)  )\cdot k_{\rho}\left(  \omega
,\rho;s\right)  =1.\label{sign}%
\end{equation}

On the other hand, the normal $n=\nabla\phi(k)$ to the surface $\Gamma(\rho)$
at $k\left(  \omega,\rho;s\right)  $ is parallel to $\omega$ with the same
direction$.$ Hence,
\[
\nabla\phi(k\left(  \omega,\rho;s\right)  )=c\omega,\quad c=c(\omega
,\rho;s)>0.
\]
From here and (\ref{sign})$,$ it follows that
\[
\omega\cdot k_{\rho}\left(  \omega,\rho;s\right)  >0.
\]
The left-hand side of this inequality is equal to $\mu_{\rho},$ since
$\mu\left(  \omega,\rho;s\right)  =k\left(  \omega,\rho;s\right)  \cdot\omega
$. Thus $\gamma_{s}=+1$, and therefore (\ref{newasym1}) leads to (\ref{asym}).

Our next step is to prove that the functions $\psi_{\pm}=R_{\lambda\pm i0}f$
satisfy (\ref{rads1}). We shall need the following estimate of the function
$\Phi$ defined in (\ref{411th})$:$%
\begin{equation}
\frac1R\underset{\xi\in B_{R}}{\sum}\left[  \left|  \Phi\left(  \xi
,\rho\right)  \right|  ^{2}+\left|  \frac\partial{\partial\rho}\Phi\left(
\xi,\rho\right)  \right|  ^{2}\right]  \leq C,\quad\left|  \rho-\lambda
\right|  \leq\delta,\quad R\geq1.\label{103}%
\end{equation}

In order to prove (\ref{103}), we cover $\Gamma\left(  \lambda\right)  $ by a
finite number of balls $B^{n},$ $n=1,...,N,$ such that the orthogonal
projection of $\Gamma\left(  \lambda\right)  \cap\overline{B^{n}}$ into one of
the hyperplanes $k_{j}=0,$ $j=j\left(  n\right)  ,$ is smooth. Then, the same
balls cover $\Gamma\left(  \rho\right)  $ with the same property of smoothness
of the projections if $\left|  \rho-\lambda\right|  \leq\delta_{2} $ and
$\delta_{2}$ is small enough. One can construct a partition of unity on
$\underset{\left|  \rho-\lambda\right|  \leq\delta}{\cup}\Gamma\left(
\rho\right)  $ subordinate to the covering $\left\{  B^{n}\right\}  ,$ i.e.,
find $N$ functions $\alpha_{n}\left(  k\right)  \in C^{\infty}\left(
T^{d}\right)  $ with support in $B^{n}$ such that $\overset{N}{\underset
{n=1}{\sum}}\alpha_{n}\left(  k\right)  =1$ on $\Gamma\left(  \rho\right)
,\left|  \rho-\lambda\right|  \leq\delta_{2}.$ Then $\Gamma\left(
\rho\right)  ,$ $\left|  \rho-\lambda\right|  \leq\delta_{2},$ has the
following form on the support of $\alpha_{n}:$%
\[
k_{j}=g\left(  k^{\prime},\rho\right)  ,\quad k^{\prime}=(k_{1},...,k_{j-1}%
,k_{j+1},...,k_{d}),\quad j=j(n).
\]
We shall choose the constant $\delta$ in (\ref{ResL2}) and (\ref{4th}) so
small that $\delta<\delta_{2}.$ Then we write the function $\Phi$ in the form
\begin{equation}
\Phi\left(  \xi,\rho\right)  =\overset{N}{\underset{n=1}{\sum}}\Phi_{n}\left(
\xi,\rho\right)  ,\quad\text{ }\Phi_{n}\left(  \xi,\rho\right)  =\left(
2\pi\right)  ^{1-d/2}\int_{\Gamma\left(  \rho\right)  }\frac{\alpha_{n}\left(
k\right)  \chi\left(  k\right)  \widehat{f}\left(  k\right)  e^{ik\cdot\xi}%
}{\left|  \nabla\phi\right|  }ds.\label{101}%
\end{equation}

Obviously,
\begin{equation}
\Phi_{n}\left(  \xi,\rho\right)  =\left(  2\pi\right)  ^{1-d/2}\int_{T^{d-1}%
}\left.  \left(  \frac{\alpha_{n}\left(  k\right)  \chi\left(
k\right) \widehat{f}\left(  k\right)  }{\left|  \nabla\phi\right|
}\sqrt{1+\left| \nabla g\right|  ^{2}}e^{ik\cdot\xi}\right)
\right| _{k_{j}=g\left(
k^{\prime},\rho\right)  }dk^{\prime},\label{102}%
\end{equation}
where $j=j(n).$

Let us estimate each term in (\ref{101}). Note that $\widehat{f}$
is analytic, since $f$ has a bounded support. Hence, the integrand
is infinitely smooth, and when differentiating (\ref{102}) with
respect to $\rho,$ the derivative can be passed under the integral
sign. Thus, Lemma \ref{lem11} implies that (\ref{103}) holds for
the functions $\Phi_{n}$, and this proves (\ref{103}) for $\Phi.$

From (\ref{103})$,$ it follows that the second term on the
right-hand side of (\ref{4th}) satisfies (\ref{rads1}). Moreover,
estimate (\ref{103}) implies (\ref{rads1}) also for the first term
on the right-hand side of (\ref{4th})$.$ Indeed, if
$F=\mathrm{Re}\Phi ,$ then
\[
\int_{\left|  \rho-\lambda\right|  \leq\delta}\frac{F\left(  \xi
,\rho\right)  }{\rho-\lambda}d\rho=\int_{\left|
\rho-\lambda\right| \leq\delta}\frac{F\left(  \xi,\rho\right)
-F\left(  \xi,\lambda\right) }{\rho-\lambda}d\rho=\int_{\left|
\rho-\lambda\right|  \leq\delta}F_{\rho }\left( \xi,\theta\left(
\rho\right)  \right)  d\rho,
\]
where $\left|  \theta\left(  \rho\right)  -\lambda\right|  \leq\left|
\rho-\lambda\right|  \leq\delta,$ and therefore, for any $R\geq1,$
\[
\underset{\xi\in B_{R}}{\sum}\left|  \int_{\left|
\rho-\lambda\right| \leq\delta}\frac{F\left(  \xi,\rho\right)
}{\rho-\lambda}d\rho\right| ^{2}\leq\underset{\xi\in
B_{R}}{\sum}2\delta\int_{\left|  \rho-\lambda\right|
\leq\delta}\left|  F_{\rho}\left(  \xi,\theta\left(  \rho\right)
\right) \right|  ^{2}d\rho\leq4\delta^{2}CR.
\]
Similar estimates are valid for $\mathrm{Im}\Phi .$ This proves
(\ref{rads1}) for the first term on the right-hand side of
(\ref{4th}). Since the remainder in (\ref{4th}) decays uniformly
in $\omega\in\Omega,$ this proves that $\psi_{\pm}=R_{\lambda\pm
i0}f$ satisfy (\ref{rads1}). Since it was already proven that
these functions satisfy (\ref{asym}), we have that $\psi_{\pm}\in
W_{\pm}.$

Finally, it remains to prove the uniqueness of the solution. So, we need to
prove that if $\psi_{\pm}\in W_{\pm}^{\prime}$ are solutions of the
homogeneous equation $\left(  \Delta-\lambda\right)  \psi=0,$ then $\psi_{\pm
}=0.$ \ Let $\xi^{0}$ be an arbitrary point of\textbf{\ }$\mathbb{Z}^{d}, $
and let $E_{\pm}$ be fundamental solutions of the difference operator
$\Delta-\lambda$ :
\[
\left(  \Delta-\lambda\right)  E_{\pm}=\delta\left(  \xi\right)  ,\quad
E_{\pm}=R_{\lambda\pm i0}\delta\left(  \xi\right)  .
\]
Let us apply the Green formula for the difference Laplacian to $\psi_{\pm}%
(\xi)$ and $E_{\pm}(\xi-\xi^{0})$ in a cube $B_{r}$ of $\mathbb{Z}^{d}$ that
contains the point $\xi^{0},$ $B_{r}=\left[  -r,r\right]  ^{d},$
$r\in\mathbb{N},$
\begin{equation}
\underset{\xi\in Br}{\sum}\left(  \Delta\psi_{\pm}\cdot E_{\pm}-\psi_{\pm
}\cdot\Delta E_{\pm}\right)  \left(  \xi\right)  =\underset{\xi\in\partial
B_{r}}{\sum}\left(  \psi_{\pm}^{\prime}E_{\pm}-\psi_{\pm}E_{\pm}^{\prime
}\right)  \left(  \xi\right) \label{green2}%
\end{equation}
where $u^{\prime}:=u\left(  \xi+e\right)  $ and $e$ is the outward unit normal
to the boundary $\partial B_{r}$ of the cube $B_{r}$ at the point $\xi$. If
$\xi\in\partial B_{r}$ is a point of intersection of several faces, then the
right-hand side of (\ref{green2}) includes terms with normals to each face
containing $\xi.$ \ In (\ref{green2})$,$ one can replace $\Delta$ by
$\Delta-\lambda.$ Hence, the left-hand side of (\ref{green2}) is equal to
$-\psi\left(  \xi^{0}\right)  .$ \ Let us take the average of both sides of
equation (\ref{green2}) over $r\in(R,2R],$ $R\in\mathbb{N},$
\[
\psi_{\pm}\left(  \xi^{0}\right)  =\frac{1}{R}\underset{\xi\in B_{2R}%
\backslash B_{R}}{\sum}\left(  \psi_{\pm}\cdot E_{\pm}^{\prime}-\psi_{\pm
}^{\prime}\cdot E_{\pm}\right)  \left(  \xi\right)  .
\]
The uniqueness follows if we prove that
\begin{equation}
\frac{1}{R}\underset{\xi\in B_{2R}\backslash B_{R}}{\sum}\left(  \psi_{\pm
}\cdot E_{\pm}^{\prime}-\psi_{\pm}^{\prime}\cdot E_{\pm}\right)  \left(
\xi\right)  \rightarrow0\quad\text{as }R\rightarrow\infty.\label{rhs11}%
\end{equation}

In order to prove (\ref{rhs11})$,$ we need to improve estimate (\ref{rads1})
for the solutions $\psi_{\pm}=R_{\lambda\pm i0}f$ of equation (\ref{door2})$.$
Let $\Omega_{0}^{\prime}$ be the following extension of the set $\Omega_{0}$
of singular points:
\[
\Omega_{0}^{\prime}=\Omega_{0}\cup\{\text{ }\omega\in\Omega;\text{ }%
k_{j}(\omega,\lambda;s)=-k_{j}(\omega,\lambda;t)\text{ for some }j,s,t\}.
\]
Let $\Omega_{\varepsilon}$ be the $\varepsilon-$neighborhood on $\Omega$ of
the set $\Omega_{0}^{\prime}$:
\[
\Omega_{\varepsilon}=\{\text{ }\omega\in\Omega;\text{ dist}(\omega,\Omega
_{0}^{\prime})<\varepsilon\},
\]
and let
\[
B(R,\varepsilon)=\{\xi\in B_{2R}\backslash B_{R};\quad\text{ }\omega=\frac
{\xi}{\left|  \xi\right|  }\in\Omega_{\varepsilon}\}.
\]
We need to show that
\begin{equation}
\frac{1}{R}\underset{\xi\in B(R,\varepsilon)}{\sum}\left|  \psi_{\pm}\left(
\xi\right)  \right|  ^{2}\rightarrow0\quad\text{as }\varepsilon+\frac{1}%
{R}\rightarrow0.\label{106}%
\end{equation}

In order to prove (\ref{106})$,$ we are going first to prove a similar
estimate for the function $\Phi$ given by (\ref{411th}) and for its
derivative:
\begin{equation}
\frac1R\underset{\xi\in B(R,\varepsilon)}{\sum}\left(  \left|  \Phi\left(
\xi,\rho\right)  \right|  ^{2}+\left|  \frac\partial{\partial\rho}\Phi\left(
\xi,\rho\right)  \right|  ^{2}\right)  \rightarrow0\quad\text{as }%
\varepsilon+\frac1R+\delta\rightarrow0.\label{107}%
\end{equation}
Relation (\ref{107}) immediately leads to (\ref{106}) if we take into account
the fact that $\psi_{\pm}$ in (\ref{ResL2}) and (\ref{4th}) do not depend on
$\delta$. Indeed, (\ref{106}) can be proved using the same arguments as in the
proof of estimate (\ref{rads1}). One needs only to refer to (\ref{107})
instead of (\ref{103}) and to keep in mind that $\delta$ is arbitrarily small.

In order to prove (\ref{107})$,$ we represent $\Phi$ as before in the form
(\ref{101}). To investigate the asymptotic behavior of the integrals in
(\ref{102}) as $\left|  \xi\right|  \rightarrow\infty,$ we look for the
stationary phase points of the integrands in (\ref{101}) and (\ref{102}) for
each fixed $\omega=\frac\xi{\left|  \xi\right|  }$ and $\rho$ such that
$\left|  \rho-\lambda\right|  \leq\delta.$ For the integral in (\ref{101})$, $
these are the points $k=\pm k\left(  \omega,\rho;s\right)  \in\Gamma\left(
\rho\right)  $ where the normal vector to $\Gamma\left(  \rho\right)  $ is
parallel to $\omega=\frac\xi{\left|  \xi\right|  }.$ Since the integral
(\ref{102}) is obtained by using local coordinates $k^{\prime}$ in
(\ref{101})$,$ the stationary phase points in (\ref{102}) are the projections
$\pm k^{\prime}\left(  \omega,\rho;s\right)  $ of the points $k\left(
\omega,\rho;s\right)  $ into the $k^{\prime}$-hyperplane.

We represent (\ref{102}) as a sum of two functions $\Phi_{n}=\Phi_{n,1}%
+\Phi_{n,2}$ with additional factors $h_{_{\alpha}}$ and $\left(  1-h_{\alpha
}\right)  ,$ respectively, in the integrand, where $h_{_{\alpha}} $ is the
following function. Let
\[
K=\{k\in T^{d};\quad k=\pm k(\omega,\lambda,s),\text{ }\omega\in\Omega
_{0}^{\prime}\},
\]
and let $K^{\prime}$ be the orthogonal projection of the set $K$ into the
torus $T^{d-1}$ defined by the equation $k_{j}=0,$ $j=j(n).$ Let $K_{\alpha
}^{\prime}$ be the $\alpha-$neighborhood (in $T^{d-1})$ of the set $K^{\prime
}.$ Since the set $K^{\prime}$ is analytic,
\[
\text{meas}K_{\alpha}^{\prime}\rightarrow0\quad\text{as }\alpha\rightarrow0.
\]
We choose $h_{_{\alpha}}=h_{_{\alpha}}(k^{\prime})$ in such a way that
$h_{_{\alpha}}\in C^{\infty}(T^{d-1}),$ $\left|  h_{\alpha}\right|  \leq1,$
$h_{\alpha}=1$ on $K_{2\alpha}^{\prime},$ and $h_{\alpha}=0$ outside of
$K_{3\alpha}^{\prime}.$ Then we apply Lemma \ref{lem11} to $\Phi_{n,1}$ and to
$\frac{\partial}{\partial\rho}\Phi_{n,1}.$ In the integral representations of
these two functions, the $L^{2}$-norm of the integrands tends to zero as
$\alpha\rightarrow0,$ due to the presence of the factor $h_{_{\alpha}}.$
Hence, if $\left|  \rho-\lambda\right|  <\delta^{\prime}$ and $\delta^{\prime
}$ is small enough then
\[
\frac{1}{R}\underset{\xi\in B_{2R}}{\sum}\left(  \left|  \Phi_{n,1}\left(
\xi,\rho\right)  \right|  ^{2}+\left|  \frac{\partial}{\partial\rho}\Phi
_{n,1}\left(  \xi,\rho\right)  \right|  ^{2}\right)  \rightarrow0\quad\text{as
}\alpha\rightarrow0.
\]
Thus, for any $\gamma>0,$ there is an $\alpha=\alpha(\gamma)$ such that, for
any $\varepsilon,$
\begin{equation}
\frac{1}{R}\underset{\xi\in B(R,\varepsilon)}{\sum}\left(  \left|
\underset{n}{\sum}\Phi_{n,1}\left(  \xi,\rho\right)  \right|  ^{2}+\left|
\frac{\partial}{\partial\rho}\underset{n}{\sum}\Phi_{n,1}\left(  \xi
,\rho\right)  \right|  ^{2}\right)  <\frac{\gamma}{4}\quad\text{as }%
\alpha<\alpha(\gamma).\label{f1}%
\end{equation}

Now we apply the stationary phase method to functions $\Phi_{n,2}$ which are
given by the integrals (\ref{102}) with an additional factor $1-$
$h_{_{\alpha}\text{ }}$ in the integrand. We take into account that the
functions $(\omega,\rho)\rightarrow k^{\prime}(\omega,\rho,s)$ are continuous,
and we choose $\varepsilon=\varepsilon(\gamma)$ and $\delta=\delta
(\gamma)<\delta^{\prime}$ to be so small that the points $\pm k^{\prime
}(\omega,\rho,s)$ belong to $K_{\alpha(\gamma)}^{\prime}$ when $\omega
\in\Omega_{\varepsilon(\gamma)},$ $\left|  \rho-\lambda\right|  <\delta
(\gamma).$ Since the stationary phase points are the points $k^{\prime}=\pm
k^{\prime}\left(  \omega,\rho;s\right)  ,$ and $1-$ $h_{_{\alpha}}=0$ in a
neighborhood of those points, we obtain that $\Phi_{n,2}$ and $\frac{\partial
}{\partial\rho}\Phi_{n,2}$ are of order $O(\left|  \xi\right|  ^{-\infty})$ as
$\left|  \xi\right|  \rightarrow\infty,$ i.e. for any $m$ and some
$C_{m}\left(  \alpha\right)  $%
\begin{equation}
\left|  \underset{n}{\sum}\Phi_{n,2}\left(  \xi,\rho\right)  \right|
^{2}+\left|  \frac{\partial}{\partial\rho}\underset{n}{\sum}\Phi_{n,2}\left(
\xi,\rho\right)  \right|  ^{2}\leq\frac{C_{m}\left(  \alpha\right)  }{\left|
\xi\right|  ^{m}},\quad\frac{\xi}{\left|  \xi\right|  }\in\Omega_{\varepsilon
},\quad\left|  \rho-\lambda\right|  \leq\delta.\label{109}%
\end{equation}

The number of points in $B(R,\varepsilon),$ $R>1,$ does not exceed
$(2R+1)^{d}<(3R)^{d}.$ Thus, (\ref{109}) with $m=d$ implies estimate
(\ref{f1}) for $\sum\Phi_{n,2}$ if
\[
\omega=\frac{\xi}{\left|  \xi\right|  }\in\Omega_{\varepsilon(\gamma)}%
,\quad\left|  \rho-\lambda\right|  <\delta(\gamma),\quad R>R(\gamma
)=C_{d}\left(  \alpha\right)  3^{d}\gamma/4.
\]
This estimate together with (\ref{f1}) proves (\ref{107}). Thus, (\ref{106})
is also proved.

Now we return to the proof of (\ref{rhs11})$.$ In fact, it is enough to prove
that
\begin{equation}
\frac{1}{R}\underset{\xi\in B_{2R}\backslash B_{R}}{\sum}\left(  \psi_{\pm
}^{\prime}\cdot E_{\pm}\right)  \left(  \xi\right)  \rightarrow0\quad\text{as
}R\rightarrow\infty,\label{1term}%
\end{equation}
since a similar relation for the second part in the sum (\ref{rhs11}) can be
proved in the same way$.$ Note that $E_{\pm}\left(  \xi-\xi^{0}\right)  \in
W_{\pm}^{\prime},$ since this inclusion was proved for $R_{\lambda\pm i0}f $
if $f$ has a bounded support, in particular, if $f=\delta(\xi-\xi^{0}).$ \ By
the same reason, (\ref{106}) is valid for $E_{\pm}\left(  \xi-\xi^{0}\right)
:$%
\begin{equation}
\frac{1}{R}\underset{\xi\in B(R,\varepsilon)}{\sum}\left|  E_{\pm}\left(
\xi-\xi^{0}\right)  \right|  ^{2}\rightarrow0\quad\text{as }\varepsilon
+\frac{1}{R}\rightarrow0.\label{door3}%
\end{equation}
\ The proof of (\ref{1term}) will be based on (\ref{door3}) and on the facts
that $\psi_{\pm}\left(  \xi\right)  $ and $E_{\pm}\left(  \xi-\xi^{0}\right)
$ are in $W_{\pm}^{\prime}.$

Let $D_{\varepsilon}=\Omega\backslash\Omega_{\varepsilon}.$
Obviously, $D_{\varepsilon}$ is a compact subset of $\Omega,$ and
$D_{\varepsilon}$ can be represented as a union of a finite number
of compacts, $D_{\varepsilon
}=\overset{m}{\underset{i=1}{\cup}}D_{\varepsilon,i},$ such that
each $D_{\varepsilon,i}$ is contained in a non-singular domain
$V_{i}\subset \Omega.$ We represent $B_{2R}\backslash B_{R}$ as a
union of $B(R,\varepsilon)$ and sets
\[
B_{i}(R,\varepsilon)=\{\xi\in B_{2R}\backslash B_{R},\quad\text{ }\omega
=\frac{\xi}{\left|  \xi\right|  }\in D_{\varepsilon,i}\},\quad i=1,...,m.
\]

Obviously,
\begin{equation}
\left|  \underset{B(R,\varepsilon)}{\sum}\left(  \psi_{\pm}^{\prime}\cdot
E_{\pm}\right)  \left(  \xi\right)  \right|  \leq\left(  \underset
{B(R,\varepsilon)}{\sum}\left|  \psi_{\pm}^{\prime}\left(  \xi\right)
\right|  ^{2}\right)  ^{\frac12}\left(  \underset{B(R,\varepsilon)}{\sum
}\left|  E_{\pm}\left(  \xi-\xi^{0}\right)  \right|  ^{2}\right)  ^{\frac
12}.\label{52a}%
\end{equation}
Note that the estimate (\ref{rads1}) holds for $\psi_{\pm}^{\prime}$ if it
holds for $\psi_{\pm}.$ From that estimate, (\ref{52a}) and (\ref{door3}) it
follows that
\begin{equation}
\left|  \frac1R\underset{\xi\in B(R,\varepsilon)}{\sum}\left(  \psi_{\pm
}^{\prime}\cdot E_{\pm}\right)  \left(  \xi\right)  \right|  \rightarrow
0\quad\text{as }\varepsilon+\frac1R\rightarrow0\text{ }.\label{52'}%
\end{equation}
Hence, (\ref{1term}) will be proved if we show that for each $\varepsilon>0$
and each $i=1,...,m,$
\begin{equation}
\left|  \frac1R\underset{\xi\in B_{i}(R,\varepsilon)}{\sum}\left(  \psi_{\pm
}^{\prime}\cdot E_{\pm}\right)  \left(  \xi\right)  \text{ }\right|
\rightarrow0\quad\text{as }R\rightarrow\infty.\label{2term}%
\end{equation}
Since $\psi_{\pm}$ satisfy (\ref{34'}) and (\ref{rads2}), we have that
$\psi_{\pm}^{\prime}$ satisfy the same relations (see arguments in the second
paragraph of the proof of Theorem \ref{asym11}). Hence,
\[
\psi_{\pm}^{\prime}\left(  \xi\right)  =\overset{m_{V_{i}}}{\underset
{s=1}{\sum}}\varphi_{s\pm}\left(  \xi\right)  \quad\text{and\quad}E_{\pm
}\left(  \xi-\xi^{0}\right)  =\overset{m_{V_{i}}}{\underset{s=1}{\sum}}%
\zeta_{s\pm}\left(  \xi\right)  ,\quad\xi\in B_{i}(R,\varepsilon),
\]
where $\varphi_{s\pm}$ and $\zeta_{s\pm}$ satisfy the conditions
(\ref{rads2})$.$ Thus, in order to prove (\ref{2term}) and to complete the
proof of Theorem\textbf{\ }\ref{asym11}, it is enough to show that for each
$\varepsilon>0$ and each $i=1,...,m,$ and $1\leq s,t\leq m_{V_{i}},$%
\begin{equation}
I:=\frac1R\underset{\xi\in B_{i}(R,\varepsilon)}{\sum}\left(
\varphi_{s\pm}\cdot\zeta_{t\pm}\right)  \left(  \xi\right)
\rightarrow0\quad\text{as }R\rightarrow\infty.\label{3ter}%
\end{equation}

When $\omega\in D_{\varepsilon,i},$ there exists a constant $c(\varepsilon)>0$
such that
\begin{equation}
\left|  \text{ }e^{\pm i[k_{j}(\omega,\lambda;s)+k_{j}(\omega,\lambda
;t)]}-1\right|  >c(\varepsilon),\quad1\leq j\leq d.\label{54'}%
\end{equation}
Now we fix $j=1$ and recall that $e_{1}=\left(  1,0,...,0\right)  .$ We
represent the terms in (\ref{3ter}) in the form
\[
\frac{e^{\pm i[k_{1}(\omega,\lambda;s)+k_{1}(\omega,\lambda;t)]}\varphi_{s\pm
}\cdot\zeta_{t\pm}}{e^{\pm i[k_{1}(\omega,\lambda;s)+ik_{1}(\omega
,\lambda;t)]}-1}-\frac{\varphi_{s\pm}\cdot\zeta_{t\pm}}{e^{\pm i[k_{1}%
(\omega,\lambda;s)+k_{1}(\omega,\lambda;t)]}-1},
\]
and then we use (\ref{rads2})$.$ This allows us to rewrite $I$ as
$R\rightarrow\infty$ in the following form:
\[
I=\frac1R\underset{\xi\in B_{i}(R,\varepsilon)}{\sum}\left[  \frac{\left(
\varphi_{s\pm}\cdot\zeta_{t\pm}\right)  \left(  \xi+e_{1}\right)  }{e^{\pm
i[k_{1}(\omega,\lambda;s)+k_{1}(\omega,\lambda;t)]}-1}-\frac{\left(
\varphi_{s\pm}\cdot\zeta_{t\pm}\right)  \left(  \xi\right)  }{e^{\pm
i[k_{1}(\omega,\lambda;s)+k_{1}(\omega,\lambda;t)]}-1}\right]  +o\left(
1\right)  .
\]
Due to (\ref{54'})$,$ we have
\[
\left|  \frac1{e^{\pm i[k_{1}(\omega,\lambda;s)+k_{1}(\omega,\lambda;t)]}%
-1}-\frac1{e^{\pm i[k_{1}(\omega^{\prime},\lambda;s)+k_{1}(\omega^{\prime
},\lambda;t)]}-1}\right|  \leq\frac CR,
\]
if $\omega^{\prime}=\frac{\xi+e_{1}}{\left|  \xi+e_{1}\right|  },$ $\xi\in
B_{i}(R,\varepsilon)$ and $R$ is big enough. Thus,
\begin{equation}
I=\frac1R\underset{\xi\in B_{i}(R,\varepsilon)}{\sum}\left[  u\left(
\xi+e_{1}\right)  -u\left(  \xi\right)  \right]  +o(1),\quad u\left(
\xi\right)  =\frac{\left(  \varphi_{s\pm}\cdot\zeta_{t\pm}\right)  \left(
\xi\right)  }{e^{\pm[ik_{1}(\omega,\lambda;s)+k_{1}(\omega,\lambda;t)]}%
-1}.\label{starv}%
\end{equation}

Let $C_{\varepsilon}$ be the cone in $\mathbb{R}^{d},$ defined by the
conditions $\frac{\xi}{\left|  \xi\right|  }$ $\in\partial\Omega_{\varepsilon
},$ $\left|  \xi\right|  >0$. Then, (\ref{starv}) implies that
\begin{equation}
\left|  I\right|  \leq\frac{1}{R}\underset{\xi\in A_{1}}{\sum}\left|  u\left(
\xi\right)  \right|  +\underset{\xi\in A_{2}}{\sum}\left|  u\left(
\xi\right)  \right|  +o(1),\label{5ter}%
\end{equation}
where $R\rightarrow\infty,$
\[
A_{1}=\left\{  \xi\in B_{2R+1}\backslash B_{R-1},\quad\text{ dist}%
(\xi,C_{\varepsilon})\leq1\right\}
\]
and
\[
A_{2}=\left\{  \xi\in\mathbb{Z}^{d},\quad\text{dist}(\xi,\partial\left(
B_{2R}\backslash B_{R}\right)  )\leq1,\quad\frac{\xi}{\left|  \xi\right|  }\in
D_{\varepsilon,i}\right\}  .
\]
Obviously, $\frac{\xi}{\left|  \xi\right|  }\in D_{\frac{\varepsilon}{2},i}$
if $\xi\in A_{1}$ and $R$ is big enough. Hence, the first relation in
(\ref{rads2}) is valid for the functions $\varphi_{s\pm}$ and $\zeta_{t\pm}$
when $\xi\in A_{1}\cup A_{2}$ and $R\rightarrow\infty.$ Thus,
\begin{equation}
\left|  u\right|  =O\left(  R^{1-d}\right) ,\quad\text{as }\xi\in
A_{1}\cup A_{2},\text{ }R\rightarrow\infty.\label{6ter}%
\end{equation}

Since the number of points in $A_{1}\cup A_{2}$ does not exceed
$CR^{d-1},$ (\ref{5ter}) and (\ref{6ter}) lead to (\ref{3ter})$.$%
\endproof
\bigskip

\noindent\textbf{3. The Equation with a Potential: }In this section, we extend
the results of the previous sections to the Schr\"{o}dinger equation $\left(
\Delta-\lambda+q\right)  \psi=f,$ where $f$ and $q$ are two functions from
$C_{0}\left(  \mathbb{Z}^{d}\right)  $ and $q$ is real valued$.$

Let the support of $q$ be contained in the cube $\left[  -r,r\right]  ^{d},$
for some positive integer $r.$ Also, let $N=\left(  2r+1\right)  ^{d}$ be the
upper bound on the number of points of the lattice in the support of $q.$

\begin{theorem}
The spectrum of $\Delta+q,$ outside the interval $\left[  -2d,2d\right]  ,$
consists of at most $N$ real eigenvalues less than $-2d$ and at most $N$ real
eigenvalues greater than $2d.$
\end{theorem}%

\proof
This statement is a direct consequence (see \cite{gla}, Theorem 13$^{bis}$ of
Chapter I) of the facts that $Sp\left(  \Delta\right)  =\left[  -2d,2d\right]
$ and that the rank of the operator of multiplication by $q$ does not exceed
$N$.%
\endproof

\begin{lemma}
\label{lq1}Let $q$ and $f$ be any two functions from $C_{0}\left(
\mathbb{Z}^{d}\right)  .$ The relation
\[
\psi=R_{_{\lambda}}\varphi,\text{ when }\operatorname{Im}\lambda\neq0\text{
}(\text{resp. }\psi_{\pm}=R_{\lambda\pm i0}\varphi,\text{ when }\lambda\in
S=\left[  -2d,2d\right]  \backslash S_{0}),
\]
gives a one-to-one correspondence between the solutions $\psi\in l^{2}\left(
\mathbb{Z}^{d}\right)  $ (resp. $\psi_{\pm}\in W_{\pm})$ of the equation
\[
\left(  \Delta-\lambda+q\right)  \psi=f,\text{ }\operatorname{Im}\lambda
\neq0\text{ }(\text{resp. }\lambda\in S),
\]
and the solutions $\varphi\in C_{0}\left(  \mathbb{Z}^{d}\right)  $ of
\[
(I+T_{_{\lambda}})\varphi=f,\text{ }(\text{resp. }(I+T_{\lambda\pm i0}%
)\varphi=f),
\]
where $T_{_{\lambda}}=qR_{_{\lambda}},$ $($resp. $T_{\lambda\pm i0}%
=qR_{\lambda\pm i0}).$
\end{lemma}%

\proof
Let $\psi\in l^{2}\left(  \mathbb{Z}^{d}\right)  $ be a solution of
\begin{equation}
\left(  \Delta-\lambda+q\right)  \psi=f,\text{ }\operatorname{Im}\lambda
\neq0.\label{im}%
\end{equation}
Then, $\left(  \Delta-\lambda\right)  \psi=f-q\psi,$ and $\psi=R_{_{\lambda}%
}\varphi$ with $\varphi=f-q\psi\in C_{0}\left(  \mathbb{Z}^{d}\right)  $ and
$R_{_{\lambda}}$ given by (\ref{res})$.$ By substituting this value of $\psi$
into (\ref{im}), we get
\begin{equation}
\varphi+qR_{_{\lambda}}\varphi=f,\text{ }\operatorname{Im}\lambda
\neq0.\label{imm}%
\end{equation}

Conversely, let $\varphi\in C_{0}\left(  \mathbb{Z}^{d}\right)  $ be a
solution of (\ref{imm}). Then, $\psi=R_{_{\lambda}}\varphi\in l^{2}\left(
\mathbb{Z}^{d}\right)  $ and satisfies the relation $\left(  \Delta
-\lambda\right)  \psi=\varphi.$ From this relation and (\ref{imm}), it follows
that (\ref{im}) holds. Thus, the statement of this lemma for
$\operatorname{Im}\lambda\neq0$ is proved.

In the case when $\lambda\in S,$ the statement can be proved in a similar way.%
\endproof

\begin{theorem}
\label{tu}Any solution of the homogeneous equation $\left(  \Delta
-\lambda+q\right)  \psi=0,$ $\lambda\in S=\left[  -2d,2d\right]  \backslash
S_{0},$ that belongs to one of the classes $W_{\pm},$ belongs also to the
other one.
\end{theorem}%

\proof
To be specific, let $\psi\in W_{+}.$ We apply the Green formula (\ref{green2})
to $\psi$ and $\overline{\psi},$ the complex conjugate of $\psi, $ in a cube
$B_{r}=\left[  -r,r\right]  ^{d}\subset\mathbb{Z}^{d},$ where $r$ is a
positive integer. Thus,
\[
\underset{\xi\in B_{r}}{\sum}\left(  \Delta\psi\cdot\overline{\psi}-\psi
\cdot\Delta\overline{\psi}\right)  \left(  \xi\right)  =\underset{\xi
\in\partial B_{r}}{\sum}\left(  \psi^{\prime}\overline{\psi}-\psi
\overline{\psi}^{\prime}\right)  \left(  \xi\right)  .
\]

Since both $\psi$ and $\overline{\psi}$ are solutions of the homogeneous
Schr\"{o}dinger equation, the left-hand side of the Green formula is equal to
zero. By taking the average with respect to $r\in(R,2R],$ we get
\begin{equation}
\frac1R\underset{\xi\in B_{2R}\backslash B_{R}}{\sum}\left(  \psi^{\prime
}\overline{\psi}-\psi\overline{\psi}^{\prime}\right)  \left(  \xi\right)
=0.\label{sumdB}%
\end{equation}

We are going to use some arguments from the proof of Theorem \ref{asym11}
where the expression similar to the left-hand side of (\ref{sumdB}) was
studied. The difference is that one of the factors in (\ref{sumdB}) contains
the complex conjugation, which is not present in (\ref{rhs11})$.$ That is why
we need to change slightly the definition of the set $\Omega_{0}^{\prime
}\subset\Omega$ in order to be able to refer to the proof of Theorem
\ref{asym11}. We denote by $\Omega_{0}^{\prime}$ the following set:
\[
\Omega_{0}\cup\left\{  \omega\in\Omega;\text{ }k_{j}\left(  \omega
,\lambda;s\right)  =\text{ }k_{j}\left(  \omega,\lambda;t\right)  \text{ for
some }j,s,t\text{ with }s\neq t\right\}  .
\]
We preserve the same definition of the sets $B(R,\varepsilon)$ and
$B_{i}(R,\varepsilon)$ introduced in the proof of Theorem \ref{asym11}, but
with the new set $\Omega_{0}^{\prime}$ in these definitions.

We split the sum (\ref{sumdB}) in two parts: over $\xi\in B(R,\varepsilon)$
and over $\xi\in$ $\underset{i}{\cup}B_{i}(R,\varepsilon).$ On the first set,
\begin{equation}
\frac1R\underset{\xi\in B(R,\varepsilon)}{\sum}\left(  \psi^{\prime}%
\overline{\psi}-\psi\overline{\psi}^{\prime}\right)  \left(  \xi\right)
\rightarrow0\quad\text{as }\varepsilon+\frac1R\rightarrow0.\label{anahid}%
\end{equation}
Indeed, from Theorem \ref{asym11} it follows that $\psi=R_{\lambda+i0}\left(
-q\psi\right)  ,$ and therefore (\ref{106}) is valid for $\psi.$ Now
(\ref{anahid}) follows similarly to (\ref{52'})$.$

Note that the expansion (\ref{asym}) is valid for $\psi:$%
\begin{equation}
\psi\left(  \xi\right)  =\text{ }\overset{m_{i}}{\underset{s=1}{\sum}}%
\frac{e^{i\mu\left(  \omega,\lambda;s\right)  \left|  \xi\right|  }}{\left|
\xi\right|  ^{\frac{d-1}2}}a\left(  \omega,\lambda;s\right)  +O\left(
\frac1{\left|  \xi\right|  ^{\frac{d+1}2}}\right)  \quad\text{as }\xi\in
V_{i},\quad\left|  \xi\right|  \longrightarrow\infty,\label{anahid1}%
\end{equation}
where $m_{i}=m_{V_{i}}.$ As it was mentioned in the proof of Theorem
\ref{asym11}, the same expansion is valid for $\psi^{\prime}.$ Thus, as
$R\rightarrow\infty,$%
\[
\frac1R\underset{\xi\in B_{i}(R,\varepsilon)}{\sum}\left(  \psi^{\prime
}\overline{\psi}-\psi\overline{\psi}^{\prime}\right)  \left(  \xi\right)
=\frac1R\underset{1\leq s\neq t\leq m_{i}}{\sum}\text{ }\underset
{B_{i}(R,\varepsilon)}{\sum}\left(  \psi_{s}^{\prime}\overline{\psi}_{t}%
-\psi_{s}\overline{\psi}_{t}^{\prime}\right)  \left(  \xi\right)
\]
\begin{equation}
+\frac1R\overset{m_{i}}{\underset{s=1}{\sum}}\text{ }\underset{B_{i}%
(R,\varepsilon)}{\sum}\frac1{\left|  \xi\right|  ^{d-1}}\left|  a\left(
\omega,\lambda;s\right)  \right|  ^{2}+o\left(  1\right)  ,\label{anahid2}%
\end{equation}
where $\psi_{s}$ are the terms under the summation sign in (\ref{anahid1}).
The first sum on the right-hand side of (\ref{anahid2}) (which includes the
functions $\psi_{s},$ $\psi_{t},$ $s\neq t)$ tends to zero as $R\rightarrow
\infty.$ This can be proved by repeating the arguments which led to
(\ref{2term}). From here, (\ref{sumdB}), (\ref{anahid}) and (\ref{anahid2}),
it follows that for any $i,$%
\[
\frac1R\overset{m_{i}}{\underset{s=1}{\sum}}\text{ }\underset{\xi\in
B_{i}(R,\varepsilon)}{\sum}\frac1{\left|  \xi\right|  ^{d-1}}\left|  a\left(
\omega,\lambda;s\right)  \right|  ^{2}\rightarrow0\quad\text{as }%
\varepsilon+\frac1R\rightarrow0.
\]
This is possible only if all $a\left(  \omega,\lambda;s\right)  $ are zeros.
In this case, $\psi$ belongs to both classes $W_{\pm}$%
\endproof

\begin{theorem}
\label{ts1}For any $\lambda\in S,$ and for any $q\in C_{0}\left(
\mathbb{Z}^{d}\right)  ,$ the homogeneous equation $\left(  \Delta
-\lambda+q\right)  \psi=0$ has only trivial solutions in the classes $W_{\pm}$.
\end{theorem}%

\proof
Let $\lambda\in S$ and
\[
\Gamma_{c}\left(  \lambda\right)  =\left\{  k\in\mathbb{C}^{d}%
/\operatorname{mod}2\pi\mathbb{Z}^{d},\text{ }\varphi\left(  k\right)
=\lambda\right\}  ,
\]
i.e., $\Gamma_{c}\left(  \lambda\right)  $ consists of points $k\in
\mathbb{C}^{d}$ such that $\varphi\left(  k\right)  =\lambda$ and with any two
points $k^{\left(  1\right)  }$, $k^{\left(  2\right)  }$ identified if
$\left(  k^{\left(  1\right)  }-k^{\left(  2\right)  }\right)  /2\pi
\in\mathbb{Z}^{d}.$ This is an analytic $\left(  d-1\right)  $-dimensional
complex manifold without singularities because%

\begin{equation}
\nabla\phi=-2\left(  \sin k_{1},...,\sin k_{d}\right)  \neq0\text{ \quad on
}\Gamma_{c}\left(  \lambda\right)  .\label{grad1}%
\end{equation}

It is not difficult to check that $\Gamma_{c}\left(  \lambda\right)  $ is
connected and its intersection with the real space $\mathbb{R}^{d}$ is
$\Gamma\left(  \lambda\right)  $ given by (\ref{star101}). Since $\Gamma
_{c}\left(  \lambda\right)  $ is connected and smooth, it is irreducible (see
section 1 of \cite{chir} or section 2 of \cite{ku}).

Let $\psi$ be a solution of $\left(  \Delta-\lambda+q\right)  \psi=0.$ Then
$\left(  \Delta-\lambda\right)  \psi=f,\quad$where $f=-q\psi\in C_{0}\left(
\mathbb{Z}^{d}\right)  .$ According to Theorem \ref{tu}, $\psi$ belongs to
both classes $W_{\pm},$ and therefore the coefficients $a_{\pm}$ in expression
(\ref{asym}) for $\psi$ are zeros. From the remark to Theorem \ref{asym11} it
follows that this can only happen if $\widehat{f}\left(  k\right)  =0$ on the
set $\Gamma\left(  \lambda\right)  .$ Note that $\widehat{f}\left(  k\right)
$ is invariant under the transformation $k\rightarrow k+2\pi\xi,$ $\xi
\in\mathbb{Z}^{d}.$ Since $\Gamma\left(  \lambda\right)  $ (which is
$\Gamma_{c}\left(  \lambda\right)  \cap\mathbb{R}^{d})$ is a $\left(
d-1\right)  $-dimensional real manifold, $\Gamma_{c}\left(  \lambda\right)  $
is irreducible and $\widehat{f}\left(  k\right)  =0$ on $\Gamma\left(
\lambda\right)  $, we have that $\widehat{f}\left(  k\right)  =0$ on
$\Gamma_{c}\left(  \lambda\right)  $ ( see \cite{ku}, Corollary 2.7). From
here and (\ref{grad1}), it follows that $\frac{\widehat{f}\left(  k\right)
}{\phi(k)-\lambda}$ is an entire function of $k\in\mathbb{C}^{d}.$

Let $z_{j}=e^{ik_{j}}$ be the conformal transformation which maps the strip
$\left|  \operatorname{Re}k_{j}\right|  <\pi$ into the complex $z_{j}$-plane
without the origin. With the new variables, $\widehat{f}$ takes the form%

\[
\widehat{f}\left(  k\right)  =\underset{\xi\in supp(f)}{\sum}f\left(
\xi\right)  e^{-ik\cdot\xi}=\underset{\xi\in supp(f)}{\sum}f\left(
\xi\right)  z^{-\xi},\text{ \quad where }z^{^{\xi}}=\overset{d}{\underset
{j=1}{\prod}}z_{j}^{\xi_{j}}.
\]
Obviously, this function can be written as
\[
\widehat{f}\left(  k\right)  =F\left(  z\right)  \overset{d}{\underset
{j=1}{\prod}}z_{j}^{-\alpha_{j}},
\]
where $F\left(  z\right)  $ is a polynomial of $z$ and the $\alpha_{j}$ are
positive integers.

Similarly,
\[
\phi(k)-\lambda=\overset{d}{\underset{i=1}{\sum}}\left(  z_{i}+\frac1{z_{i}%
}\right)  -\lambda=Q\left(  z\right)  \overset{d}{\underset{i=1}{\prod}}%
z_{i}^{-1},
\]
where%

\[
Q\left(  z\right)  =\left(  \overset{d}{\underset{i=1}{\sum}}z_{i}\right)
\overset{d}{\underset{i=1}{\prod}}z_{i}+\overset{d}{\underset{j=1}{\sum}%
}\left(  \overset{d}{\underset{i\neq j}{\prod}}z_{i}\right)  -\lambda
\overset{d}{\underset{i=1}{\prod}}z_{i}.
\]

Then,
\begin{equation}
\frac{\widehat{f}\left(  k\right)  }{\phi(k)-\lambda}=\frac{F\left(  z\right)
}{Q\left(  z\right)  }\overset{d}{\underset{i=1}{\prod}}z_{i}^{1-\alpha_{i}%
}.\label{eqz1}%
\end{equation}
The right-hand side of $\left(  \ref{eqz1}\right)  $ is analytic for all
complex $z$ except possibly on the hyperplanes $z_{j}=0,$ $j=1,...,d$ (where
the transform: $k\rightarrow z$ is not defined). Thus, the same is true for
$\frac{F\left(  z\right)  }{Q\left(  z\right)  }.$ But $Q\left(  z\right)  $
is not equal to zero if $z_{j_{0}}=0$ for some $j_{0},$ and $z_{j}\neq0$ for
$j\neq j_{0}.$ Hence, $\frac{F\left(  z\right)  }{Q\left(  z\right)  }$ is
analytic except possibly on some set of complex dimension $d-2$ (the
intersection of two hyperplanes). Therefore, $\frac{F\left(  z\right)
}{Q\left(  z\right)  }$ is an entire function (see \cite{kra}, Corollary 7.3.2).

Since $F\left(  z\right)  $ and $Q\left(  z\right)  $ are both polynomials and
$\frac{F\left(  z\right)  }{Q\left(  z\right)  }$ is an entire function, then
$\frac{F\left(  z\right)  }{Q\left(  z\right)  }$ is a polynomial (see
\cite{sha}, appendix, a corollary to The Hilbert Nullstellensatz). Hence from
(\ref{eqz1}), it follows that $\widehat{\psi}\left(  k\right)  =\frac
{\widehat{f}\left(  k\right)  }{\phi(k)-\lambda}$ is a trigonometric
polynomial, which implies that $\psi\left(  \xi\right)  $ has bounded support.
The only solution of the homogeneous equation with a bounded support is
$\psi\equiv0.$%
\endproof

Let us recall that we denote the resolvent $\left(  \Delta-\lambda+q\right)
^{-1}$ by $\mathcal{R}_{_{\lambda}}$ if $q\neq0$ and denote it by
$R_{_{\lambda}}$ if $q=0.$

\begin{theorem}
For any $f\in$ $C_{0}\left(  \mathbb{Z}^{d}\right)  ,$ any $q\in$
$C_{0}\left(  \mathbb{Z}^{d}\right)  ,$ and any $\lambda\in S,$ the limits
$\psi_{\pm}$ of $\psi_{_{\eta}}=\mathcal{R}_{_{\eta}}f$ as $\eta
\rightarrow\lambda\pm i0$ exist$. $ Moreover, for each $\lambda\in S,$ the
equation
\begin{equation}
\left(  \Delta-\lambda+q\right)  \psi=f\label{starv1}%
\end{equation}
admits unique solutions in $W_{+}$ and in $W_{-},$ and these solutions are
$\psi_{\pm}=\mathcal{R}_{\lambda\pm i0}f.$
\end{theorem}%

\proof
The homogeneous equation $\left(  \Delta-\eta+q\right)  \psi=0$ has only
trivial solutions in $l^{2}\left(  \mathbb{Z}^{d}\right)  $ if
$\operatorname{Im}\eta\neq0.$ Therefore, by Lemma \ref{lq1}, $I+T_{_{\eta}}$
has only a trivial kernel in this case. Thus, if $\operatorname{Im}\eta\neq0,$
$I+T_{_{\eta}}$ is invertible since $T_{_{\eta}}$ is a finite-dimensional operator.

Similarly, from Theorem \ref{ts1} and Lemma \ref{lq1}, it follows that
$I+T_{\lambda\pm i0}$ is invertible if $\lambda\in S.$ Theorem \ref{at}
implies that $T_{_{\eta}}=qR_{_{\eta}}\rightarrow qR_{\lambda\pm
i0}=T_{\lambda\pm i0}$ as $\eta\rightarrow\lambda\pm i0$ if $\lambda\in S.$
Thus, for $\lambda\in S,$
\begin{equation}
\left(  I+T_{_{\eta}}\right)  ^{-1}\rightarrow\left(  I+T_{\lambda\pm
i0}\right)  ^{-1}\quad\text{as }\operatorname{Im}\eta\neq0,\text{ }%
\eta\rightarrow\lambda\pm i0\text{ }.\label{lim10}%
\end{equation}

Moreover, we have that $\mathcal{R}_{_{\eta}}=R_{_{\eta}}\left(  I+T_{_{\eta}%
}\right)  ^{-1}.$ By applying the limiting absorption principle to $R_{_{\eta
}}$ and using (\ref{lim10}), we get, for $\lambda\in S,$ that
\begin{equation}
\mathcal{R}_{_{\eta}}\rightarrow R_{\lambda\pm i0}\left(  I+T_{\lambda\pm
i0}\right)  ^{-1}\quad\text{as }\operatorname{Im}\eta\neq0,\text{ }%
\eta\rightarrow\lambda\pm i0,\label{74'}%
\end{equation}
and this concludes the proof of the first statement of the theorem.

From (\ref{74'}), it also follows that $\mathcal{R}_{\lambda\pm i0}%
f=R_{\lambda\pm i0}\varphi,$ where $\varphi=\left(  I+T_{\lambda\pm
i0}\right)  ^{-1}f\in C_{0}\left(  \mathbb{Z}^{d}\right)  .$ Thus, the
solutions $\psi_{\pm}=\mathcal{R}_{\lambda\pm i0}f$ of equation (\ref{starv1})
belongs to $W_{\pm}$ due to Theorem \ref{asym11}. The uniqueness of solutions
in $W_{\pm}$ is proved in Theorem \ref{ts1}.%
\endproof
\bigskip

\noindent

\noindent\textbf{5. Appendix: }

\medskip

\noindent\textbf{Proof of Lemma \ref{convexity}.} We shall assume that
$\lambda>0$ since the case $\lambda<0$ can be studied similarly. At any point
$k\in\Gamma(\lambda),$ the curvature of the surface is equal to
\[
K(k)=\frac{\overset{d}{\underset{j=1}{\sum}}\sin^{2}k_{j}\overset{d}%
{\underset{\underset{m\neq j}{m=1}}{\Pi}}\cos k_{m}}{\left(  \overset
{d}{\underset{j=1}{\sum}}\sin^{2}k_{j}\right)  ^{\frac{d+1}{2}}},\quad
k=\left(  k_{1},...,k_{d}\right)  .
\]
For the sake of simplicity of formulas we shall provide the proof of the Lemma
only in the cases $d=2$ and $d=3$. Let
\[
s_{i}=\overset{i}{\underset{j=1}{\sum}}\cos k_{j}\quad\text{and\quad}%
p_{i}=\overset{i}{\underset{\underset{}{j=1}}{\Pi}}\cos k_{j}.
\]

When $d=2,$ the equation of the surface $\Gamma(\lambda)$ is
\[
\cos k_{1}+\cos k_{2}=\frac{\lambda}{2},\quad\text{where }\left(  k_{1}%
,k_{2}\right)  \in\left[  -\pi,\pi\right]  ^{2}.
\]
The total curvature $K(k)$ of this surface at any point $k=\left(  k_{1}%
,k_{2}\right)  $ is equal to
\[
K(k)=\frac{\cos k_{1}\sin^{2}k_{2}+\sin^{2}k_{1}\cos k_{2}}{\left(  \sin
^{2}k_{1}+\sin^{2}k_{2}\right)  ^{\frac{3}{2}}}=\frac{s_{2}\left(
1-p_{2}\right)  }{\left(  \sin^{2}k_{1}+\sin^{2}k_{2}\right)  ^{\frac{3}{2}}%
}>0,
\]
since $\lambda\in\left(  0,4\right)  ,$ $s_{2}=\lambda/2>0,$ and $0\leq\left|
p_{2}\right|  <1.$ The convexity of $\Gamma(\lambda)$ follows from the facts
that $K(k)>0$ and that $\Gamma(\lambda)$ is located strictly inside the square
$\left[  -\pi,\pi\right]  ^{2}$( inside $\left[  0,2\pi\right]  ^{2}$ if
$\lambda<0).$

When $d=3,$ the spectrum of the difference Laplacian is $Sp(\Delta)=[-6,6],$
and the total curvature $K(k)$ of the surface $\Gamma(\lambda)$ at any point
$k$ is equal to
\begin{equation}
K(k)=\frac{\cos k_{1}\cos k_{2}\sin^{2}k_{3}+\cos k_{1}\cos k_{3}\sin^{2}%
k_{2}+\cos k_{2}\cos k_{3}\sin^{2}k_{1}}{\left(  \sin^{2}k_{1}+\sin^{2}%
k_{2}+\sin^{2}k_{3}\right)  ^{2}}.\label{ana}%
\end{equation}

If $\cos k_{j}>0$ for $1\leq j\leq3,$ then $K(k)>0$ due to (\ref{ana}). Else,
only one of the three cosines is negative because otherwise $s_{3}<1,$ and
that contradicts the assumption of the Lemma that $\lambda\in\left(
2,6\right)  $ if $\lambda>0.$ Without loss of generality, let $\cos k_{3}<0.$
It is easy to show that the curvature can be written in the following form:
\begin{equation}
K(k)=\frac{p_{3}}{\left(  \sin^{2}k_{1}+\sin^{2}k_{2}+\sin^{2}k_{3}\right)
^{2}}\left(  \frac1{\cos k_{3}}+\frac{s_{2}}{p_{2}}-s_{3}\right)
.\label{ana2}%
\end{equation}
Since $s_{3}>1,$
\begin{equation}
\cos k_{3}>1-s_{2}.\label{ana3}%
\end{equation}
Moreover,
\begin{equation}
p_{2}\geq s_{2}-1,\label{ana4}%
\end{equation}
due to the fact that $\cos k_{1,2}\leq1.$ By combining the inequality $s_{3}>1
$ with (\ref{ana3}) and (\ref{ana4}), we get that the last factor on the
right-hand side of (\ref{ana2}) is negative:
\[
\frac1{\cos k_{3}}+\frac{s_{2}}{p_{2}}-s_{3}<\frac1{1-s_{2}}+\frac{s_{2}%
}{s_{2}-1}-1=0.
\]
After taking into account that $p_{3}$ is negative, (\ref{ana2}) implies that
$K(k)>0$. The convexity of $\Gamma(\lambda)$ follows from the fact that
$\Gamma(\lambda)$ is located strictly inside the cube $\left[  -\pi
,\pi\right]  ^{3}.$%
\endproof
\medskip

\noindent\textbf{Proof of Lemma \ref{normals}. }Let $\omega=\left(  \omega
_{1},\omega_{2},...,\omega_{d}\right)  $ be a unit vector and $k=\left(
k_{1},k_{2},...,k_{d}\right)  $ be a point of $\Gamma(\lambda)$ that satisfies
the assumptions of the Lemma$.$ Hence,
\[
\overset{d}{\underset{j=1}{\sum}}\cos k_{j}=\lambda/2\text{\qquad and\qquad
}\sin k_{j}=-\kappa\omega_{j},\quad j=1,...,d,
\]
where $\kappa$ is a positive constant$.$ Thus, $\overset{d}{\underset
{j=1}{\sum}}\left(  \pm\sqrt{1-\kappa^{2}\omega_{j}^{2}}\right)  =\frac
\lambda2.$ Since $\kappa>0,$ the number $m$ does not exceed the number of
solutions of the equations
\[
\overset{d}{\underset{j=1}{\sum}}\left(  \pm\sqrt{1-x\omega_{j}^{2}}\right)
=\frac\lambda2.
\]

Without loss of generality, we consider only $0\leq\lambda\leq2d$ and
$0\leq\omega_{1}\leq\omega_{2}\leq...\leq\omega_{d}.$ Let $f(x)$ denote one of
the following functions:
\begin{equation}
f(x)=\overset{d}{\underset{j=1}{\sum}}\left(  \pm\sqrt{1-x\omega_{j}^{2}%
}\right)  -\frac{\lambda}{2}.\label{funf}%
\end{equation}
Since there are $2^{d}$ different functions $f(x),$ $m\leq m_{0}2^{d},$ where
$m_{0}$ is the maximal number of zeros that any individual function $f(x)$ can admit$.$

We are going to prove the following assertion: For each $i=0,1,...,d,$ there
exists a function
\begin{equation}
f_{i}\left(  x\right)  =\overset{d-i}{\underset{j=1}{\sum}}a_{j}\left(
1-xb_{j}\right)  ^{\frac12-i}+c,\quad x<\frac1{b_{d-i}},\label{fi}%
\end{equation}
such that $0<b_{1}<...<b_{d-i}$ and $m_{0}\leq m_{i}+i,$ where $m_{i}$ is the
number of zeros of the function $f_{i}\left(  x\right)  .$ Let us prove this
assertion by induction.

For $i=0,$ one only needs to show that the function $f(x),$ defined in
(\ref{funf}), can be written in the form (\ref{fi}) with $i=0.$ If
$0<\omega_{1}<...<\omega_{d},$ then $f(x)$ already has the form (\ref{fi})
with $a_{j}=\pm1,b_{j}=\omega_{j}^{2},$ and $c=-\frac{\lambda}{2}$. Else if
$\omega_{1}=\omega_{2}=...=\omega_{k}=0,$ then we drop the first $k$ terms in
(\ref{funf}) and choose $c=\overset{k}{\underset{j=1}{\sum}}\left(
\pm1\right)  -\frac{\lambda}{2}.$ Also, we combine the terms in (\ref{funf})
with equal $\omega_{j}$'s$.$ Last, in order to write (\ref{funf}) in the form
(\ref{fi}), it is left to add several terms to (\ref{funf}), with small
positive $b_{j}$ and zero coefficient $a_{j},$ to keep the number of terms in
the sum (\ref{funf}) equal to $d.$

Now, suppose that the assertion holds for $i=n.$ Let us prove it for $i=n+1.$
Obviously,
\begin{equation}
f_{n}^{\prime}\left(  x\right)  =\overset{d-n}{\underset{j=1}{\sum}}%
a_{j}\left(  n-\frac{1}{2}\right)  b_{j}\left(  1-xb_{j}\right)
^{-(n+\frac{1}{2})},\quad x<\frac{1}{b_{d-n}},\label{fn'}%
\end{equation}
and $m_{n}\leq m_{n}^{\prime}+1,$ where $m_{n}^{\prime}$ is the number of
zeros of the function $f_{n}^{\prime}\left(  x\right)  .$ On the other hand,
$m_{0}\leq m_{n}+n$ because the assertion holds for $i=n.$ Thus,
\begin{equation}
m_{0}\leq m_{n}^{\prime}+n+1.\label{m0}%
\end{equation}
From (\ref{fn'}), it follows that
\[
\frac{1}{\left(  n-\frac{1}{2}\right)  \left(  1-xb_{d-n}\right)
^{-(n+\frac{1}{2})}}f_{n}^{\prime}\left(  x\right)  =\overset{d-(n+1)}%
{\underset{j=1}{\sum}}a_{j}b_{j}\left(  \frac{1-xb_{j}}{1-xb_{d-n}}\right)
^{-(n+\frac{1}{2})}+a_{d-n}b_{d-n},
\]
where $x<\frac{1}{b_{d-n}}.$ With the one-to-one change of variable
$x=\frac{1}{z}+\frac{1}{b_{d-n}},$ we get
\begin{equation}
\frac{\left(  -z\right)  ^{-(n+\frac{1}{2})}}{\left(  n-\frac{1}{2}\right)
}f_{n}^{\prime}\left(  x(z)\right)  =\overset{d-(n+1)}{\underset{j=1}{\sum}%
}a_{j}b_{j}^{\frac{1}{2}-n}\left(  1-z(\frac{1}{b_{j}}-\frac{1}{b_{d-n}%
}\right)  ^{-(n+\frac{1}{2})}+a_{d-n}b_{d-n}^{\frac{1}{2}-n},\label{fi1}%
\end{equation}
where $z<0.$ Let $f_{n+1}\left(  z\right)  $ be the function in the left-hand
side of this equation$.$ The number of negative zeros of this function is
equal to $m_{n}^{\prime}.$

Certainly, the function $f_{n+1}\left(  z\right)  $ will have the form
(\ref{fi}) if one changes the index of summation in (\ref{fi1}): $j\rightarrow
d-n-j.$ Moreover, this function is defined for $z<\frac{1}{b_{1}}-\frac
{1}{b_{d-n}},$ and the number $m_{n+1}$ of its zeros on this bigger interval
is not less than the number $m_{n}^{\prime}$ of its negative zeros. This,
together with (\ref{m0}), proves the assertion for $i=n+1.$ Hence by
induction, the assertion is true for any $i=0,...,d.$

If $a_{j}\neq0$ for any $j=1,...,d,$ in particular when the $\omega_{j}$'s are
distinct and positive, the function $f_{d}\left(  x\right)  $ is constant and
$f_{d}\left(  x\right)  =a_{1}b_{1}^{-d+\frac32}\neq0.$ In this case,
$m_{d}=0$ and $m_{0}\leq m_{d}+d=d.$ Based on the construction of the
functions $f_{i}\left(  x\right)  ,$ one can check that if some of the $a_{j}%
$'s are zero, one of the functions $f_{i}\left(  x\right)  $ is a non zero
constant. For this $i,$ $m_{0}\leq m_{i}+i=i\leq d.$ So, $m_{0}\leq d,$ and
$m\leq m_{0}2^{d}\leq2^{d}d.$%
\endproof


\begin{thebibliography}{99}
\bibitem{atk}F. V. Atkinson : \textit{Discrete and continuous boundary
problems}, Academic Press, New York and London 1964.

\bibitem{blo}E. D. Bloch : \textit{A first course in geometric topology and
differential geometry}, Birkh\"{a}user, Boston 1997.

\bibitem{chir}E. M. Chirka : \textit{Complex analytic sets, Encyclopedia of
mathematical sciences}, vol. 7, Several complex variables I,
Springer-Verlag, Berlin 1990.

\bibitem{es}M. S. Eskina: The scattering problem for partial-difference
equations, in \textit{Mathematical Physics}, Naukova Dumka, Kiev,
1967 (in Russian), 248-273.

\bibitem{es2}M. S. Eskina: The direct and the inverse scattering problem
for a partial-difference equation, \textit{Soviet Math. Doklady},
v.7, no.1 (1966), 193-197

\bibitem{gla}I. M. Glazman : \textit{Direct methods of qualitative spectral
analysis of singular differential operators, Israel Program for
scientific translation}, Jerusalem 1965 and by Daniel Davey and
Co., New York 1966.

\bibitem{kra}S. G. Krantz : \textit{Function theory of several complex
variables}, John Wiley \& Sons Inc., 1982.

\bibitem{ku}P. Kuchment and B. Vainberg : On absence of embedded
eigenvalues for Schr\"{o}dinger operators with perturbed periodic
potentials, \textit{Comm. PDE}, 25, no. 9-10 (2000), 1809 - 1826.

\bibitem{mol}S. Molchanov : \textit{Lectures on random media, Lecture notes
in mathematics}, vol. 1581, Springer Verlag, Berlin 1994.

\bibitem{sha}I. R. Shafarevich : \textit{Basic algebraic geometry: Varieties
in projective space}, vol. 1, Springer-Verlag, Berlin Heidelberg
and New York 1944.

\bibitem{ts}A. N. Tikhonov and A. A. Samarskii: \textit{Equations of
mathematical physics}, Pergamon Press Ltd, Canada 1963.

\bibitem{v}B. R. Vainberg: \textit{Asymptotic methods in equations of
mathematical physics}, Gordon and Breach, New York 1989.
\end{thebibliography}
\end{document}